\documentclass[11pt]{article}
\usepackage{amsfonts}
\usepackage{amsmath,amssymb}
\usepackage{bbm}
\usepackage{bm}
\usepackage{color}
\usepackage{slashed}
\usepackage{cite}
\usepackage[utf8]{inputenc}
\usepackage{graphicx}
\usepackage{bbold}
\DeclareGraphicsExtensions{.pdf,.png,.jpg}

\usepackage[colorlinks=false,citecolor=cyan,urlcolor=blue,bookmarks=true,bookmarks=true,bookmarksopen=true,bookmarksnumbered=true,bookmarksopenlevel=3]{hyperref}

\definecolor{airforceblue}{rgb}{0.36, 0.54, 0.66}
\definecolor{steelblue}{rgb}{0.27, 0.51, 0.71}
\definecolor{amber}{rgb}{1.0, 0.49, 0.0}

\allowdisplaybreaks[1]
\def\simg{{\ \lower-1.2pt\vbox{\hbox{\rlap{$>$}\lower6pt\vbox{\hbox{$\sim$}}}}\ }}
\def\siml{{\ \lower-1.2pt\vbox{\hbox{\rlap{$<$}\lower6pt\vbox{\hbox{$\sim$}}}}\ }}

\makeatletter \@addtoreset{equation}{section} \makeatother

\begin{document}

\flushbottom

\begin{titlepage}

\begin{centering}

\vfill

{\Large{\bf
Scalar dark matter coannihilating with a coloured fermion}
}

\vspace{0.8cm}

S.~Biondini$^{\rm a}$\footnote{ s.biondini@rug.nl}
and S.~Vogl$^{\rm b}$\footnote{ stefan.vogl@mpi-hd.mpg.de}

\vspace{0.8cm}

{\em $^{\rm{a}}$Van Swinderen Institute, University of Groningen
 \\ 
Nijenborgh 4, NL-9747 AG Groningen, Netherlands} 
\\
\vspace{0.15 cm}
{\em $^{\rm{b}}$Max-Planck-Institut 
f\"ur Kernphysik (MPIK),\\ Saupfercheckweg 1, 69117 Heidelberg,
Germany}
\vspace*{0.8cm}

\end{centering}
\vspace*{0.3cm}
 
\noindent

\textbf{Abstract}: We analyse the phenomenology of a simplified model for a real scalar dark matter candidate interacting with quarks via a coloured fermionic mediator. 
In the coannihilation regime, the dark matter abundance is controlled by the dynamics of the coloured fermions which can be significantly affected by non-perturbative effects. We employ a non-relativistic effective field theory approach which allows us to systematically treat the Sommerfeld effect and bound-state formation in the early Universe. 
The parameter space compatible with the dark matter relic abundance is confronted with direct, indirect and collider searches. A substantial part of the parameter space, with dark matter masses up to 18 TeV, is already excluded by XENON1T. Most of the remaining  thermal relics can be probed by a future Darwin-like experiment,  when taking properly into account the running of the relevant couplings for the direct detection processes.

\vfill

\vfill
\newpage
\tableofcontents
\setcounter{page}{1}
\end{titlepage}
\setcounter{page}{2}
\setcounter{footnote}{0}
\section{Introduction}
Even though evidence for dark matter (DM) has been established only through gravitational observations, it is widely accepted that an additional matter component, besides ordinary matter, populates our Universe. Indeed, there are compelling measurements from astrophysics and cosmology from a broad range of scales: from the size of a galaxy to the whole Universe.
DM can nicely accommodate the  velocity distributions of stars in galaxies, as well as the large scale structure patterns on cosmological scales. The measurements of the temperature anisotropies of the Cosmic Microwave Background (CMB) by the Planck satellite allow the extraction of the DM abundance, $\Omega_{\hbox{\tiny DM}}h^2=0.1200 \pm 0.0012$ \cite{Aghanim:2018eyx}, at percent level accuracy. This constitutes the most precise determination of the DM abundance to date. 

However, there is an almost total lack of information on the nature of DM. Event though this is not the only viable option, it is reasonable that DM comes in the form of a particle. Due to very few hints on the DM properties, a plethora of models have been proposed that provide a DM candidate. Naturally, the parameters of a given particle physics model have to comply with the observed relic abundance. Therefore, the DM mass (and particle masses in the dark sector), the couplings in the dark sector and with the Standard Model (SM) can be constrained.
In order to scrutinise the viable scenarios and the experimental capabilities, it is useful to adopt a \textit{simplified model} approach for DM searches~\cite{Abdallah:2015ter,DeSimone:2016fbz}. In such a  framework, rather than considering a fully-fledged theory, bounds and constraints are set on a simpler model that captures the most relevant features. 
A DM candidate in the Weakly Interacting Massive Particle category, i.e.~DM produced by thermal freeze-out, is easily realised in this setup.
 The presence of the additional massive state can drastically change the thermal freeze-out when the coannihilating partner is close in mass with the DM particle \cite{Griest:1990kh,Edsjo:1997bg}. The accompanying state strengthens the coupling between the DM and SM bath and dilutes the DM population beyond the naive expectation. Consequently, one has to track its (co)annihilations as well. However, in the case where the coannihilation partner feels strong interactions, the annihilation cross section is not well described by the standard techniques. Most notably, Sommerfeld enhancement and bound-state effects due to repeated gluon exchange have recently been shown to have a large impact on the relic density\cite{vonHarling:2014kha,Ibarra:2015nca,Petraki:2015hla,Pierce:2017suq,Liew:2016hqo,Mitridate:2017izz,Biondini:2018pwp,Harz:2018csl,Biondini:2018ovz}.  Interestingly, bound-state formation is much more efficient at low temperatures and the thermally averaged annihilation cross section exhibits a strong and unusual temperature dependence. This keeps annihilations effective until well after the first deviation from chemical equilibrium. Consequently, the parameter space favoured from cosmology changes, and the experimental expectations have to be adjusted accordingly. The DM scalar and the accompanying coloured mediator exhibit a rich phenomenology at colliders and can also be tested at direct detection experiments.

We include the Sommerfeld and bound-state effects in medium in the derivation of the DM abundance. The impact of these contributions is captured by averaged Sommerfeld factors, that are extracted from the spectral function of the annihilating coloured fermion pairs in a thermal environment. The spectral function contains information on both scattering and bound states. The latter are sensitive to thermal scales and we take their dynamical and temperature-dependent dissociation/formation  through a plasma-modified Schr\"odinger equation into account. It is important to notice that more than one bound state can contribute in the annihilations depending on the temperature, and the spectral function method allows to include them when they are indeed formed~\cite{Kim:2016kxt,Kim:2016zyy,Biondini:2017ufr,Biondini:2018pwp}. The thermal potentials and interaction rates are derived in the framework of tailored non-relativistic effective field theories. The advantage of this approach is two-fold: we work with the suitable degrees of freedom at the energy scale of interest, namely the binding-energy/kinetic energy; the potentials and rates are obtained as matching coefficients of a potential NREFT (pNREFT). This allows for controlled approximations at finite temperature.

We use the results of the relic density calculation as input to  a study of the phenomenology. The most relevant experimental signature are multijet plus missing energy events at the LHC and DM nucleon scattering in direct detection experiments.
 
The structure of this paper is as follows. First, we introduce our simplified model and comment on the connection to other models in Sec.~\ref{Sec:Model}. Next,  we turn to the computation of the relic density in Sec.~\ref{sec:Relic_density}. After a brief comment on the form of the involved Boltzmann equation, we present the formalism for the determination of the annihilation rates in terms of spectral functions and elucidate the connection with thermal potentials and scattering rates. In Sec.~\ref{sec:Pheno} the phenomenology of the model is analysed. We pay close attention to LHC searches and direct detection  and investigate the impact of current experimental results on the parameter space preferred by thermal freeze-out. Finally, we summarise our conclusions and comment on possible improvements in Sec.~\ref{sec:Conclusions}. 
Additional information about the thermal potentials and the renormalization group equations (RGEs) in NREFT is provided in the Appendix.

\section{Simplified Model}
\label{Sec:Model}
The simplified model we consider consists of
a gauge singlet real scalar DM ($S$) and a vector-like fermion 
($F$). The latter 
is a triplet under QCD and matches the hypercharge of the SM fermion it couples to. However, the hypercharge coupling of the fermion plays a marginal role since its effects are subleading compared with QCD effects and we omit it in the following.
The Lagrangian for this model can be expressed as 
\begin{eqnarray}
 \mathcal{L} & = & 
 \mathcal{L}^{ }_{\hbox{\tiny SM}}  + \frac{1}{2} \partial_\mu S \partial^\mu S
 - \frac{M_S^2}{2}\, S^2 
 - \frac{\lambda^{ }_2}{4!} S^4 -  \frac{\lambda^{ }_3}{2}\, S^2 \, H^\dagger H  \nonumber 
 \\
 &+& \bar{F} \left(  i \slashed{D} - M_F \right)  F
  - y\,  S \bar{F} P_R q 
 - y^* S \bar{q} P_L F\
 \;,
 \label{Lag_RT}
\end{eqnarray}
where $H$ is the  SM Higgs doublet, $D_\mu=\partial_\mu + i g_s A^{a}_\mu T^a$ is the QCD covariant derivative, $\lambda_3$ and $y$ are the additional scalar and Yukawa couplings corresponding to the Higgs and fermion portal while $P_R$ and $P_L$ are the right- and left-handed projectors, respectively. The DM self interaction coupling, which does not have an impact in this study, is denoted $\lambda_2$, whereas $\lambda_1$ is left for the Higgs self interaction. 

It is interesting to note that for non-zero $\lambda_3$ and small $y$ and/or large $M_F$ this model emulates scalar DM through the Higgs portal, one of the most minimal extensions of the SM that can account for DM \cite{Silveira:1985rk,McDonald:1993ex}. As the relic density is intimately connected with the direct detection cross section in the Higgs portal model, it is under increasing pressure due to the recent null-results of direct DM searches and the correct relic density can only be achieved in special regions of the parameter space \cite{Cline:2013gha,Athron:2017kgt}. Therefore, the model under consideration here can, in certain regions of parameter space,  be seen as a ``mediator-assisted" Higgs portal model.

\section{Relic density}
\label{sec:Relic_density}
The relic density stands as the main observable that any compelling DM model has to comply with. The cosmological abundance is accurately determined by measuring the CMB anisotropies and it amounts to $\Omega_{\hbox{\tiny DM}} h^2=0.1200(12)$\cite{Akrami:2018vks}. 
Upon exploiting a mechanism to produce DM particles in the early Universe, one can use the relic density as a powerful constraint on the model parameters.

We consider DM production via thermal freeze-out in the early Universe. In this scenario, the interactions keep the DM in thermal equilibrium with the plasma at high temperatures. The corresponding processes are  pair creation and annihilation. When the temperature of the expanding Universe drops below the DM mass ($T \siml M_S $), the  processes get Boltzmann suppressed and the annihilation cannot keep up with the expansion of the Universe described by the Hubble rate $H$. A larger annihilation cross section leads to a smaller DM abundance because the heavy particles remain in equilibrium  longer and track their exponentially suppressed equilibrium number density. 

In models with coannihilating partners, thermal freeze-out gets modified by the presence of other dark sector particles in the thermal bath during freeze-out.  The efficiency of the coannihilation processes depends strongly on the mass splitting between the DM particle and the coannihilating specie. For $\Delta M /M_S \siml 0.2 $ with $\Delta M \equiv M_F - M_S $, the coloured partner has a non-negligible population compared to DM particles and can affect the abundance of the latter. This is mainly due to the efficient conversion rates between the $S$ and $F$ species that put them in thermal contact until late times (see Sec.~\ref{sec:CONV_rate}).

The effect of coloured fermions close in mass with the DM can be captured by a single Boltzmann equation 
\cite{Gondolo:1990dk,Griest:1990kh}
\begin{equation}
\frac{dn}{dt} + 3 Hn =-\langle \sigma_{{\rm{eff}}} v \rangle (n^2-n^2_{{\rm{eq}}}) \, ,
\label{BE_gen}
\end{equation}
where $H$ is the Hubble rate of the expanding Universe and $n$ denotes the overall number density of dark sector states $S$ and $F$.
Then, the total equilibrium number density, which accounts for both the particle species of the dark sector ($S$ and $F$), is
\begin{equation}
n_{{\rm{eq}}}= \int_{\bm{p}} e^{-E_{p}/T} \left[ 1 + 4 N_c \, e^{-\Delta M/T} \right] \, , 
\label{n_eq}
\end{equation}
 where $\int_{\bm{p}} \equiv \int d^3 \bm{p}/(2\pi)^3 $, $E_p=M_S+p^2/(2 M_S)$,   $p \equiv |\bm{p}|$ and the effective thermally averaged annihilation cross section reads 
\begin{equation}
\langle \sigma_{{\rm{eff}}} v \rangle = \sum_{i,j} \frac{n^{\hbox{\scriptsize eq}}_i \,  n^{\hbox{\scriptsize eq}}_j}{(\sum_k n_k^{\hbox{\scriptsize eq}})^2} \langle \sigma_{ij} v  \rangle \,.
\label{co_cross}
\end{equation}
This quantity includes all the combinations for the annihilating pairs, namely $SS$, $S F$, $F \bar{F}$, $F F$ and their conjugates when relevant. Conventionally, it is calculated by thermally averaging the in-vacuum cross sections over the centre-of-mass energies in the thermal environment.  However, this way of extracting $\langle \sigma_{{\rm{eff}}} v \rangle$ has to be treated with caution if DM and/or coannihilating partner have interactions with light plasma constituents. In this case, non-relativistic particle pairs can be severely affected by non-perturbative dynamics induced through a repeated exchange of the light particles, i.e. Sommerfeld enhancement for scattering states and bound-state formation. The latter effect has been only recently addressed in DM models.  Various formalisms are in use in the literature~\cite{vonHarling:2014kha,Liew:2016hqo,Mitridate:2017izz,Biondini:2018pwp,Harz:2018csl,Biondini:2018ovz}. The main common outcome of these studies is that larger DM masses are compatible with the relic density because a larger cross section is found. Bound-state formation is especially efficient at temperatures smaller than the initial chemical decoupling temperature and it helps to boost the late stage annihilation cross section to large values. As a consequence, the thermally averaged cross section develops a non-trivial dependence on the temperature even for s-wave annihilations.

In this paper we follow an effective field theory approach that enables to factorise the hard annihilations, occurring at the large energy scale $M_S$ ($M_F$), from the soft physics that characterises the dynamics of DM and fermion pairs in a thermal environment~\cite{Kim:2016zyy,Kim:2016kxt}: momentum transfer, kinetic energy and thermal scales. At the core of the method is the extraction of the spectral function of the annihilating non-relativistic pair, which helps us in capturing the nature of the annihilating states in a thermal plasma.

\subsection{Non-relativistic annihilations and spectral functions}
An EFT approach for Majorana fermion DM and coloured scalars has been discussed in detail in refs~\cite{Biondini:2018pwp,Biondini:2018ovz}. Here, we recall the main points and set the framework for the derivation of the thermally averaged cross section for the model under consideration. Within the freeze-out mechanism, the DM and the coannihilating partner remain in chemical equilibrium with the plasma until  $z=M_S/T \approx 25$.  
Annihilations continue even during later stages when the annihilation cross section increases due to Sommerfeld enhancement and bound-state formation. 
In this situation, most of the energy of 
DM and coloured fermions is stored in their mass. For non-relativistic species, the typical momentum
is $p= \sqrt{M_S T} = M_S \sqrt{T /M_S}$ and one can define an average velocity $v \equiv \sqrt{T /M_S}$, which is smaller than unity in the regime of interest. We take the DM mass as a reference for the large mass scale. The same applies for the coloured mediator that has a mass close to $M_S$ in the setting of the present work. Therefore, the degrees of freedom during freeze-out and later-stage annihilations are non-relativistic scalars and fermions with $M_S, M_F \gg
T$ and light SM particles.\footnote{The SM particles are taken to be light with respect to the dark particle sector. This approximation is not fully justified for the lowest DM masses considered in this work, i.e.~$M_S=500$ GeV, where the top-quark mass is only one third of the DM mass.
} 

This physical picture calls for a non-relativistic description, where energy excitations of order $M_S$ are highly suppressed. In the EFT language, hard energy/momentum modes of order $M_S$ are integrated out from the fundamental theory (\ref{Lag_RT}), whereas the smaller energy scales, both non-relativistic and thermal scales, remain dynamical.
It is worth mentioning that this first step is insensitive to plasma effects. Because of the
large energy release in the hard annihilation process, the typical distance scales are much smaller
than those introduced by the thermal plasma, i.e. $\Delta x_{\hbox{\scriptsize ann}} \approx 1/M_S \ll 1/T$.  
\begin{figure}[t!]
\centering
\includegraphics[scale=0.55]{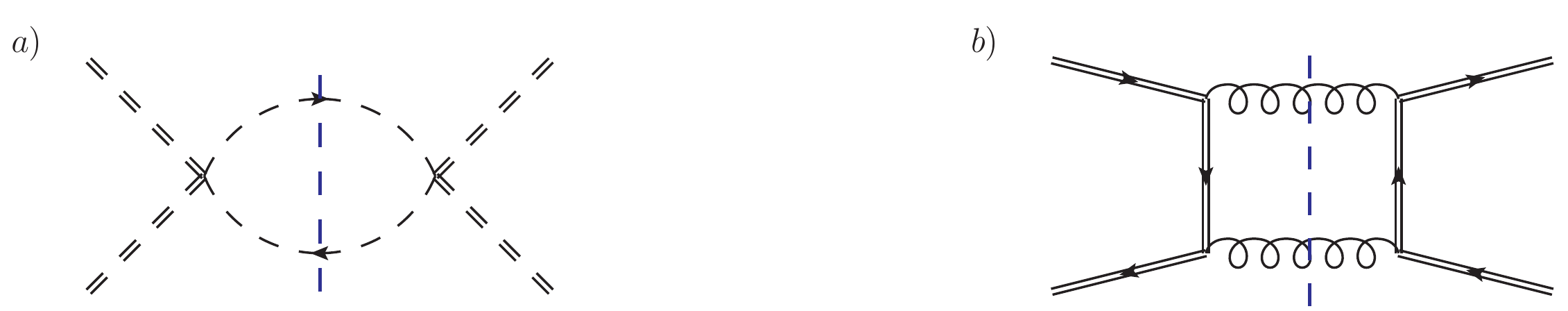}
\caption{\label{fig:Example_ann_FS}One-loop diagrams for DM pair (double-dashed lines) annihilations into SM Higgs bosons (dashed lines) and $F \bar{F}$ (double-solid lines) annihilations into a gluon pair (curly lines). The blue dashed line implements a cut along the light particles into which the heavy ones annihilate.}
\end{figure}

 In a non-relativistic EFT (NREFT) the number of heavy particles is conserved since light particles with momentum of order $M$ are not part of the theory. However, heavy particle annihilations can be described by exploiting the optical theorem.  The four-particle operators develop an imaginary part in their matching coefficients at one-loop, which encode the description of DM and coloured fermion (co)annihilations. 
 In the relativistic theory, one-loop diagrams with four-external heavy particle legs are cut along the highly energetic internal lines, see Fig.~\ref{fig:Example_ann_FS} for two representative examples. These processes, together with many others in this model, contribute to the imaginary part of the matching coefficients of the four-particle operators. They are the remnants of the relativistic processes happening at the hard scale that has been removed from the NREFT\cite{Bodwin:1994jh,Braaten:1996ix}.

In this section, we deal with heavy pair annihilations, either DM particles and coloured fermions. Therefore, we only write the Lagrangian terms that are relevant to this aim which are dimension-6 operators.  More details on the complete NREFT for this model is given in the Appx.~\ref{app:RGE}. We express the effective Lagrangian in terms of the non-relativistic scalar and fermion fields, $\phi$, $\psi$ and $\chi$. The latter two carry a spinor and colour index.  Schematically the Lagrangian reads
\begin{eqnarray}
\mathcal{L}^{\hbox{\scriptsize 4-particle}}_{\hbox{\tiny NREFT}}&=&   \mathcal{L}_{\phi\phi} + \mathcal{L}_{\phi \psi} + \mathcal{L}_{\phi \chi} + \mathcal{L}_{\psi \psi} + \mathcal{L}_{\chi \chi}  \, ,
\end{eqnarray}
and the individual terms are given as follows\footnote{In order to obtain the four-particles operators, we integrate out the mass scale $M_S$. Therefore, the residual mass splitting is still a dynamical scale. However, this is a small scale according to our assumption here, $\Delta M \ll M_S$, and we set it to zero like any other SM particle mass for the extraction of the matching coefficients.}
\begin{eqnarray} 
&&\mathcal{L}_{\phi\phi}= \frac{c_1}{M_S^2} \phi^\dagger \phi^\dagger \phi \phi \, ,
\label{SS_ann}\\ 
 \phantom{s} \nonumber 
 \\
&&\mathcal{L}_{\phi \psi} + \mathcal{L}_{\phi \chi}=  \frac{c_2}{M_S^2} ( \phi^\dagger \psi_s^{\dagger \alpha} \psi^\alpha_s \phi  + \phi^\dagger \chi_s^{\alpha} \chi_s^{ \dagger \alpha} \phi) \, ,
\label{SF_coann}
 \\ 
 \phantom{s} \nonumber 
 \\
&&\mathcal{L}_{\psi \chi}= \frac{c_3}{M_S^2} \psi^{\dagger \alpha}_s \chi_s^\alpha \,  \chi^{\dagger \beta}_r \psi_r^\beta + \frac{c_4}{M_S^2} \psi^{\dagger \alpha}_s \, (\sigma^k)_{sr} \, \chi_r^\alpha \,  \chi^{\dagger \beta}_p \, (\sigma^k)_{pq} \,  \psi_q^\beta  \nonumber
\\ 
 &&\phantom{ssss}+ \frac{c_5}{M_S^2}\psi^{\dagger \alpha}_s \, (T^a)^{\alpha \beta}\,  \chi_s^\beta \,  \chi^{\dagger \lambda}_r \, (T^a)^{\lambda \tau}\, \psi_r^\tau  + \frac{c_6}{M_S^2}   \psi^{\dagger \alpha}_s \, (\sigma^k)_{sr}  (T^a)^{\alpha \beta} \, \chi_r^\beta \,  \chi^{\dagger \lambda}_p \, (\sigma^k)_{pq} (T^a)^{\lambda \tau} \,  \psi_q^\tau   \nonumber
\\ 
 &&\phantom{ssss} + \frac{c_7}{M_S^2} \psi^{\dagger \alpha}_s \chi_r^\beta \,  \chi^{\dagger \beta}_r \psi_s^\alpha + \frac{c_8}{M_S^2} \psi^{\dagger \alpha}_s \, (\sigma^k)_{sr} \, \psi_r^\alpha \,  \chi^{\dagger \beta}_p \, (\sigma^k)_{pq} \,  \chi_q^\beta \, ,
 \label{FFbar_ann}
 \\ 
 \phantom{s} \nonumber 
 \\
&&\mathcal{L}_{\psi \psi} + \mathcal{L}_{\chi \chi}= \frac{c_9}{M_S^2}   \psi^{\dagger \alpha}_r \psi_s^{\dagger \beta} \,  \psi^{ \beta}_s \psi_r^\alpha  +  \frac{c_{10}}{M_S^2} \psi^{\dagger \alpha}_s \, (\sigma^k)_{sr} \, \psi_r^\alpha \,  \psi^{\dagger \beta}_p \, (\sigma^k)_{pq} \,  \psi_q^\beta + (\psi \to \chi^\dagger ) \, ,
\label{FF_ann}
\end{eqnarray}  
\begin{figure}[t!]
\centering
\includegraphics[scale=0.55]{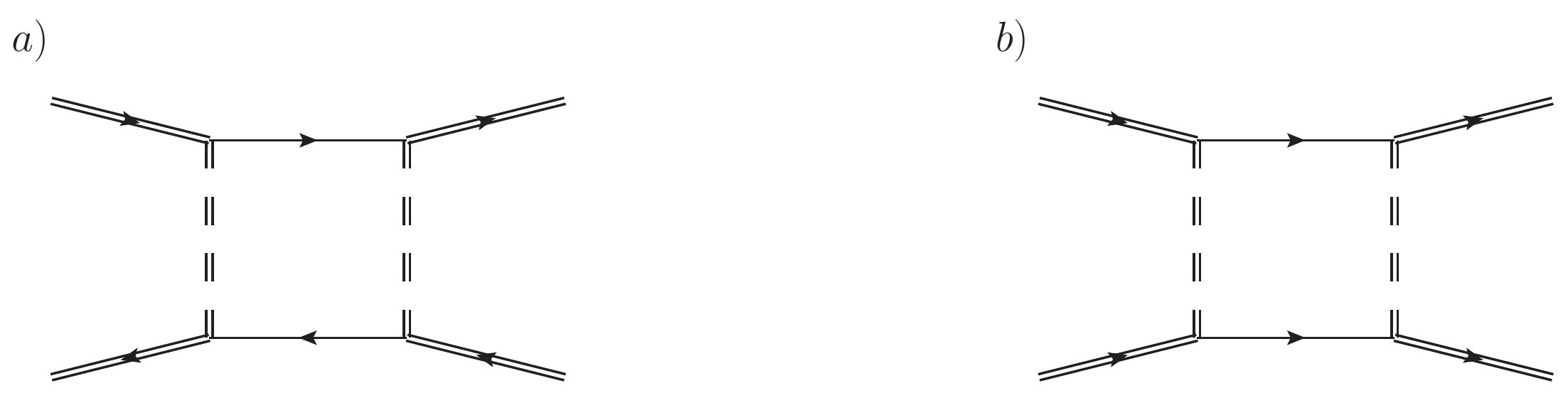}
\caption{\label{fig:FF_fig0} Additional diagrams for $F \bar{F}$ and $FF$ annihilations induced by the Yukawa interaction. Solid double lines stand for $F$, solid thin line for the SM quark and double dashed line for the scalar DM. The colour flux is different from the diagrams induced by QCD.}
\end{figure}
where $\sigma^k$ are the Pauli matrices. Some comments are in order. Scalar DM annihilations are described by the single operator in Eq.~(\ref{SS_ann}), $SF$ and $S\bar{F}$ coannihilations are comprised in Eq.~(\ref{SF_coann}), while $F \bar{F}$, $FF$ and $\bar{F}\bar{F}$ annihilations correspond to Eqs.~(\ref{FFbar_ann}), (\ref{FF_ann}) respectively. 
As far as the coloured fermion sector is concerned, the four-particle operators for $F \bar{F}$ annihilations resemble those in NRQCD \cite{Bodwin:1994jh}.\footnote{The first four operators in Eq.~(\ref{FFbar_ann})  correspond exactly to $\mathcal{O}_1(^1S_0), \mathcal{O}_1(^3 S_1), \mathcal{O}_8(^1S_0), \mathcal{O}_8(^3 S_1) $ in ref.\cite{Bodwin:1994jh} respectively, where the notation is $^{2S+1}L_J$.} 
A couple of differences occur though. First, the third line in Eq.~(\ref{FFbar_ann}) is absent in NRQCD because there is no scalar-mediated annihilations into quarks in QCD. 
Second, particle-particle (antiparticle-antiparticle) annihilations are possible due to real scalar exchange leading to $FF \to qq$ ($\bar{F}\bar{F} \to \bar{q}\bar{q}$) as shown in Fig.~\ref{fig:FF_fig0} right diagram.
It is worth mentioning that a NREFT has been considered for two heavy quarks in ref.~\cite{Brambilla:2005yk}, where two different quark species have been considered. Accordingly two independent heavy-quark fields are involved and one finds the four independent operators indicated in ref.~\cite{Brambilla:2005yk}. 
In our case, there is only one additional coloured mediator $F$, that is decomposed in the non-relativistic fields describing the heavy particle and antiparticle, $\psi$ and $\chi$ respectively. As, we have only one ``heavy-quark'' field the corresponding number of independent operators, that describe particle-particle annihilations, turns out to be two instead of four.\footnote{ One may check this by  imposing $Q=Q'$ in Eq.~(2) of ref.~\cite{Brambilla:2005yk} and using the Fierz identity for the SU(2) and SU(3) generators.
We chose the two operators that do not involve the SU(3) generators $T^a$ to write the non-relativistic Lagrangian for heavy particle-particle (antiparticle-antiparticle) annihilation in Eq.~(\ref{FF_ann}).}

At leading order in the couplings and with vanishing SM particle masses, we find the following matching coefficients 
\begin{eqnarray}
&&c_1= \frac{\lambda_3^2}{256 \pi} \, , \hspace{1.5 cm} c_2= \frac{|y|^2(|h_q|^2 + C_F  g_s^2) }{128 \pi }  \, , 
\label{c_1_2}
\\
&&  c_3 = \frac{C_F g_s^4}{32 \pi N_c} \, ,  \hspace{1.2 cm} c_4 = 0   \, ,
\\
&&   c_5 = \frac{(N_c^2-4)g_s^4}{64 \pi N_c} \, , \hspace{0.4 cm}  c_6 = \frac{N_f g_s^4}{96 \pi} + \frac{g_s^2 |y|^2}{192 \pi} \, , \\
&& c_7 = \frac{|y|^4}{256 \pi } \, ,  \hspace{1.5 cm} c_8 = \frac{|y|^4}{768 \pi } \, , \\
&& c_9= \frac{|y|^4}{256 \pi } \, ,  \hspace{1.5 cm} c_{10}= -\frac{|y|^4}{768 \pi } \, ,
\end{eqnarray}
where $N_c=3$ is the number of colours, $N_f$ the number of SM quarks and $h_q$ is the SM Yukawa coupling of the SM quarks with the Higgs boson. 

 A comment is in order for the process $SS \to q \bar{q}$. This process would be expected to contribute to coefficient $c_1$ in Eq.~(\ref{c_1_2}) but its s-wave rate is helicity suppressed. Therefore, we find a vanishing contribution to the leading four-particle operator in Eq.~(\ref{SS_ann}) when setting a vanishing quark mass. However, we aim at treating the boundary between the coannihilation regime and the standard freeze-out correctly.  This implies that we need an accurate expression for $SS$ annihilations and, especially when $\lambda_3$ is small, the scalar annihilations into quarks might be more relevant than a naive estimate suggests. The situation is different for light and top quarks. In the former case, the quark mass is so small that we have to consider the leading velocity suppressed contribution, which appears in a d-wave~\cite{Toma:2013bka,Giacchino:2013bta}. We checked that the corresponding cross section introduces a correction smaller than one per mill on the relic density for the parameter space considered in this work. Therefore, we do not consider it and avoid the computation of $1/M_S$ suppressed operators.   On the other hand, the inclusion of the top mass lifts the helicity suppression efficiently and induces a sizeable effect. Working at leading order in $m_t$ the matching coefficient of the operator in Eq.~(\ref{SS_ann}) gives $c_1^{\hbox{\tiny top}}=|y|^4 N_c (m_t/M_S)^2/(64 \pi)$ in addition to the result in Eq.~(\ref{c_1_2}). The matching coefficients parametrise the in-vacuum cross section for heavy particles annihilations.
\begin{figure}[t!]
    \centering
    \includegraphics[scale=0.55]{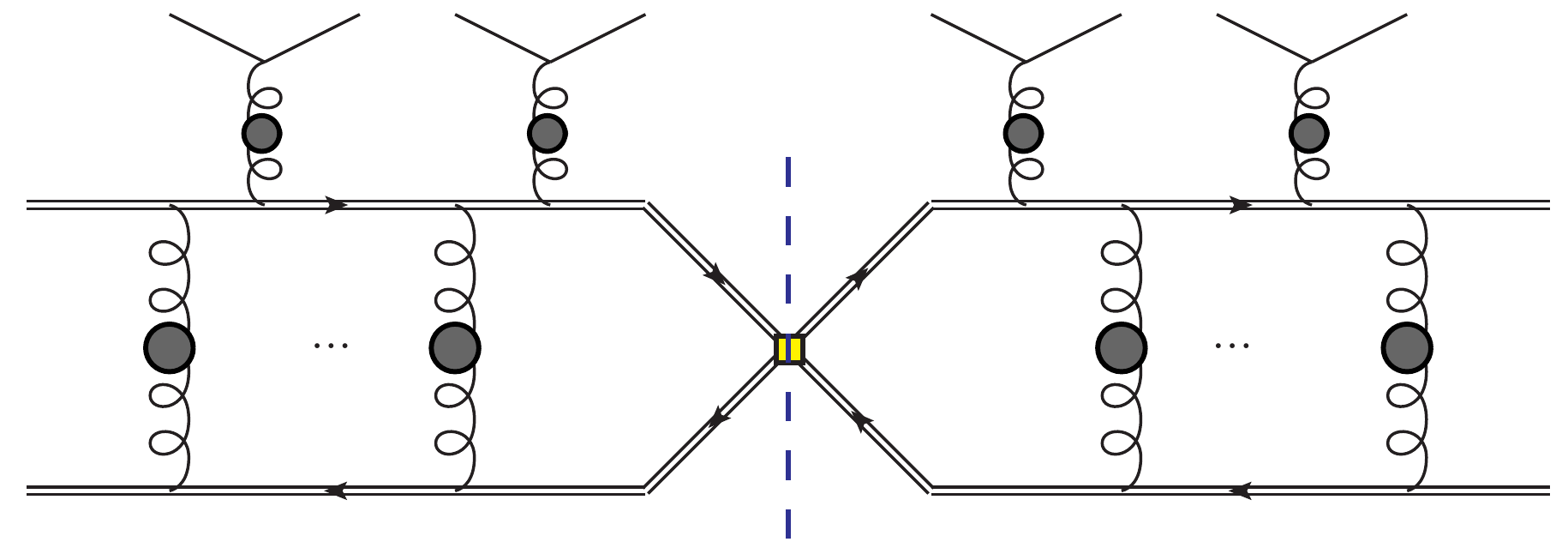}
    \caption{The diagram represent the thermal average of four-particle operators for the heavy coloured fermion in Eq.~\ref{the_aver_operator}. Repeated gluon exchange, represented by the dots, are accompanied by $2 \to 2$ scatterings with partons of the medium. These processes are characterised by the small energy scales $M_F v$, $M_Fv^2$, $\pi T$ and $m_D$. The hard annihilation is instead shown with the yellow vertex, which amounts at the matching coefficients that develop an imaginary part. This is implemented by the cut on the diagram with a blue dashed line.}
    \label{fig:thermal_cross}
\end{figure}

In addition, there is a rich set of phenomena that arise from the dynamics induced by the smaller energy scales, namely the non-relativistic momentum and kinetic energy, $M_F v$ and $M_F v^2$, and the thermal scales, namely the temperature and Debye mass for gluons. More importantly, the slow-moving coloured fermions feel QCD interactions and repeated gluon exchange can affect the heavy-pair dynamics. This leads to Sommerfeld enhancement and bound-state formation, here in a thermal environment. Moreover, the heavy particles also experience thermal mass shift and $2 \to 2$ soft scatterings with light particles of the plasma. Such processes are further accompanied by thermal gluons hitting the bound states and possibly break. These effects can be factored out from the hard annihilations and one can write the thermally averaged annihilation cross section, for which a diagrammatic representation is given in Fig.~\ref{fig:thermal_cross}, as follows~\cite{Kim:2016zyy,Kim:2016kxt}
\begin{eqnarray}
\langle \sigma_{\hbox{\scriptsize eff}} \, v \rangle &=& 2 \frac{\langle {\rm{Im}} \,  \mathcal{L}^{\hbox{\scriptsize 4-particle}}_{\hbox{\tiny NREFT}} \rangle}{n_{\hbox{\scriptsize eq}}^2} \nonumber \\
&=& \frac{2}{M_S^2 (1+4 N_c e^{-\Delta M_T /T})^2} \left\lbrace  c_1  +8 N_c \,  c_2 e^{-\Delta M_T /T} \right.  \nonumber
\\
&+& \left.   \left[ 4 N_c c_3 \, \bar{S}_s  + 2(N_c^2-1) (c_5+3 c_6) \bar{S}_o + 4 N_c^2 c_7 +  8 N_c^2 c_9 \left( \frac{1}{3} \bar{S}_{{\rm{T}}} + \frac{2}{3} \bar{S}_\Sigma \right)\right]  e^{-2 \Delta M_T /T} \right\rbrace \, ,  
\nonumber\\
\label{the_aver_operator}
\end{eqnarray}
where $n_{\hbox{\scriptsize eq}}$ has been defined in Eq.~(\ref{n_eq}) and we use $c_{10}=-c_9/3$ to obtain a more compact form for the particle-particle sector.  $c_8$ does not appear because the corresponding operator has a vanishing number of contractions. The thermally averaged Sommerfeld factors $\bar{S}_i$ are defined for the singlet, octet, antitriplet and sextet channels as follows~\cite{Kim:2016kxt,Kim:2016zyy,Biondini:2018pwp}
\begin{equation}
    \bar{S}_i = \frac{\int_{-\Lambda}^\infty \frac{d E'}{\pi} e^{-E'/T}\rho_i(E')}{\int_{-\Lambda}^\infty \frac{d E'}{\pi} e^{-E'/T}\rho^{(0)}_i(E')} = \left( \frac{4 \pi}{M_S T} \right)^{\frac{3}{2}}  \int_{-\Lambda}^\infty \frac{d E'}{\pi} e^{[{\rm{Re}} V_i(\infty)-E']/T} \frac{\rho_i(E')}{N_i} \, .
    \label{eq:sigmav}
\end{equation}
Here, $E'$ is  the energy of relative motion of the pair after factoring out the centre-of-mass dynamics, $\rho_i$  is the spectral function of the annihilating interacting pair, $\rho_i^{(0)}$ corresponds to the pair without gluon exchange but with the thermal mass shift included, and $N_i$ are the number of contractions of the operators in the Eqs.~(\ref{SS_ann})-(\ref{FF_ann}). The mass splitting accounts now for both in-vacuum and thermal contribution\footnote{ The allowed interactions in a simplified model Lagrangian are different when the coloured state is a fermion instead of a scalar. Consequently, the  contribution from the Higgs boson and gluon tadpoles to the thermal mass found in \cite{Biondini:2018pwp} do not appear here.} 
\begin{equation}
\Delta M_T = \Delta M - \frac{g_s^2 C_F m_D}{8 \pi} \, ,
\label{thermal_mass}
\end{equation}
and it originates from the real part of the self-energy of the heavy fermion with a hard thermal loop (HTL) resummed gluon propagator, whose Debye mass at leading order amounts to
\begin{equation}
    m_D=g_s T\sqrt{\frac{N_c}{3}+\frac{N_f}{6}} \, ,
\end{equation}
where $N_f$ is the number of SM quark flavours. 
The (anti)symmetrization of the source in the case of identical particles, i.e. for $FF$ ($\bar{F}\bar{F}$) annihilations, requires care. We obtain the decomposition in terms of antitriplet and sextet that have a weight of 1/3 and 2/3 respectively in agreement with refs.\cite{deSimone:2014pda,Giacchino:2015hvk}.  

The spectral functions contain  the relevant dynamical information of the annihilating heavy pair and are extracted by solving a plasma-modified Schr\"odinger equations with thermal potentials, $V_i(r,T)$, and thermal widths, $\Gamma_i(r,T)$
\begin{eqnarray}
&& \biggl[ 
   H_i(r,T) -i\Gamma_i(r,T) - E'
 \biggr] G^{ }_i(E';\bm{r},\bm{r}')  =  
 N^{ }_i\, \delta^{(3)}(\bm{r}-\bm{r}') \, ,
   \label{SH_like_1} 
 \\
 \phantom{s} \nonumber
 \\
&&\lim_{\bm{r},\bm{r}' \to \bm{0}} {\rm{Im}} \,  G^{ }_i(E';\bm{r},\bm{r}')
  =  \rho^{ }_i(E')  \, ,
  \label{SH_like}
\end{eqnarray} 
where the Hamiltonian reads $H_i(r,T)=-\nabla^2_{\bm{r}}/M_S +V_i(r,T)$ with $r=|\bm{r}|$ being the relative distance between the coloured fermions. The in-medium effects on the heavy pair dynamics are captured by the virtual and real scatterings with the plasma constituents. The latter are comprised in thermal widths, or interaction rates, that have been associated to the Landau damping and gluo-dissociation \cite{Burnier:2007qm,Brambilla:2008cx,Brambilla:2011sg,Brambilla:2013dpa}. On the other hand, the thermal potentials deviate from the in-vacuum Coulomb-like ones. The gluons acquire a thermal mass and at temperatures $\pi T \gg M_F v$ a Yukawa screened potential is established. At  smaller temperatures, $M_F v \simg \pi T $, the Coulomb potential gets thermal corrections that can be expanded in  $r T$ \cite{Brambilla:2008cx}. We stress that both the potentials and the interaction rates are function of the temperature and, therefore, an explicit derivation is possible upon assuming some relation between the non-relativistic and thermal scales. Moreover, one strength of this approach is that the composition of the annihilating pair spectrum, both scattering and bound states, is derived in a dynamical way.  

In this model, the DM particles can experience an attractive potential due to the SM Higgs. At temperatures smaller than the electroweak symmetry breaking, the SM Higgs acquires a non-vanishing expectation value $v_h$, and a vertex of the type $SSh$ is induced and one can construct ladder diagrams possibly leading to Sommerfeld enhancement and bound states.  When implementing the non-relativistic expansion for the heavy field $S$, a  vertex is generated with an effective coupling $\alpha_{\hbox{\scriptsize h}} \equiv (\lambda_3 v_h/ 2 M_S)^2/(4 \pi) \leq 0.01$~\cite{Biondini:2018xor},  for the scalar coupling value $\lambda_3 \leq 1.5$ and for the DM masses  considered here. Therefore, it is of order of weak interactions strength and we neglect the effect in our study. 
Higgs exchange and the corresponding bound-state effect has been recently considered in refs.~\cite{Harz:2017dlj,Biondini:2018xor,Harz:2019rro} for the spin-conjugate version of the simplified model addressed in this paper. Moreover, we neglect the Sommerfeld effects from exchange of the hypercharge field $B_\mu$. The corresponding effects are much smaller than those induced by the gluon exchange since they are proportional to the SM weak coupling.

\subsection{Thermal potentials and pNREFTs}
\label{sec:thermal_potentials_NREFT}
The potentials between the coloured fermions can be interpreted as matching coefficients of a potential NREFT (pNREFT)\cite{Pineda:1997bj,Brambilla:1999xf,Brambilla:2008cx}. Here the analogy with pNRQCD is rather manifest, since the accompanying vector-like fermion is a colour triplet of SU(3) and it closely resembles a heavy SM quark. Strictly speaking, a potential emerges when the typical momentum transfer between the pair, namely $M_Fv$, is integrated out 
and one is left with the kinetic/binding energy $M_Fv^2$ as the dynamical scale of the theory. This is exactly the energy scale we want to address. In the EFT language, any energy scale larger than the kinetic/binding energy can be encoded in the potential between a heavy coloured pair irrespective of being an in-vacuum or a thermal scale. 
Making contact with the formalism developed for heavy quarkonium in a thermal environment~\cite{Brambilla:2008cx,Brambilla:2010cs,Brambilla:2010vq,Brambilla:2011sg}, we shall write the pNREFT for this model that will enable us to identify the relevant degrees of freedom; allow for a rigorous derivation of the thermal potentials; include the relevant processes for bound-state formation/dissociation in a unified framework. 

In extracting the spectral functions in Eqs.~(\ref{SH_like_1}) and (\ref{SH_like}), we shall make use of the literature of heavy quarkonium in medium (see e.g.~\cite{Brambilla:2010cs} for a review). However, we are also going to provide two new results: the real part of the potential induced by a thermal gluon propagator in the case $\pi T \approx M_F v$ and the calculation of the antitriplet-to-sextet transition at finite temperature. The latter is the analogue of singlet-to-octet transitions derived in ref.\cite{Brambilla:2008cx}.
Indeed, in the model considered here, there are four different colour combinations of the non-relativistic pairs that are relevant. First, a colour singlet and octet from a $F\bar{F}$ pair, i.e.~$3 \otimes \bar{3} = 1 \oplus 8$. Moreover, a colour antitriplet and sextet originate from a $FF$ pair, in terms of the fundamental representations  $3 \otimes 3 = \bar{3} \oplus 6$. 

One can work with the pNRQCD Lagrangian whenever the momentum transfer $M_F v$ is integrated out, namely when we look at energy scales smaller then the typical inverse distance of the pair, therefore $M_Fv \approx 1/r$.
Scattering and bound states have a different typical velocity.
For scattering states
the average velocity can be estimated $v\sim \sqrt{T/M_F}$. In contrast, for bound states in a Coulomb potential $v \sim \alpha_s$, since the kinetic energy of the pair is of the order of the potential energy. We shall use this notation in Sec.~\ref{sec:real_part_gluo},  \ref{sec:gluodissociation} and \ref{sec:num_potential}.
However, one should be careful when thermal scales are present as well. A first classification can be envisaged based on the relative position of the temperature and $M_Fv$ scale. When $\pi T \gg M_Fv$, we are in the so-called screening regime where the gluons and light quarks get the corresponding HTL expressions. Accordingly, the potentials between the heavy pair is Debye screened and also a thermal width is generated which can be ascribed to Landau damping. In the opposite regime $\pi T \ll M_Fv$, the leading order potential between the pairs is a Coulomb potential. Nevertheless, thermal corrections affect it and a thermal width is generated, which corresponds to the gluodissociation process (a chromoelectric transition from a bound to an unbound state). We refer to original literature and reviews on the subject for a more detailed discussion \cite{Burnier:2007qm,Brambilla:2008cx, Brambilla:2010vq, Brambilla:2010cs}. 

That said, the pNRQCD Lagrangian is written in terms of the two-particle heavy fields, gluons and light quarks that carry energies smaller than $M_Fv$. We split the $F\bar{F}$ and $FF$ sector as follows
\begin{equation}
\mathcal{L}_{\hbox{\tiny pNRQCD}}=\mathcal{L}_{\hbox{\scriptsize light}} + \mathcal{L}^{F\bar{F}}_{\hbox{\tiny pNRQCD}} +  \mathcal{L}^{FF}_{\hbox{\tiny pNRQCD}} \, ,    
\end{equation}
where 
\begin{equation}
  \mathcal{L}_{\hbox{\scriptsize light}}=- \frac{1}{4} F_{\mu \nu}^a F^{a, \mu \nu} + \sum_i^{N_f} \bar{q}_i \slashed{D} q_i \, , 
\end{equation}
and the latter has to be understood as HTL resummed if the temperature has been integrated out before the scale $M_Fv$. Then the Lagrangian in the static limit may be written as follows \cite{Brambilla:2008cx} 
\begin{eqnarray}
\mathcal{L}^{F\bar{F}}_{\hbox{\tiny pNRQCD}}&=&\int d^3 r {\rm{Tr}} \left\lbrace {\rm{S}}^{\dagger} \left[ i\partial_0 - \mathcal{V}_s -\delta M_s\right] {\rm{S}} + {\rm{O}}^{\dagger} \left[ iD_0 - \mathcal{V}_o -\delta M_o\right] {\rm{O}} \right\rbrace \nonumber \\
&+& \int d^3 r \left[ {\rm{Tr}} \left\lbrace  {\rm{O}}^\dagger \bm{r} \cdot g_s \bm{E} \, {\rm{S}} + {\rm{S}}^\dagger \bm{r} \cdot g_s \bm{E} \, {\rm{O}} \right\rbrace + \frac{1}{2} {\rm{Tr}} \left\lbrace  {\rm{O}}^\dagger \bm{r} \cdot g_s \bm{E} \, {\rm{O}} + {\rm{O}}^\dagger \bm{r} \cdot g_s \bm{E} \, {\rm{O}} \right\rbrace \right] + \cdots \nonumber \\
\phantom{xx}
\label{pNRQCD_FbarF}
\end{eqnarray}
where ${\rm{S}}=S \bm{1}_c  / \sqrt{N_c}$ and ${\rm{O}}=O^a T^a / \sqrt{T_F}$ are the particle-antiparticle singlet and octet fields, $N_c=3$ the number of colours, $T_F=1/2$, $\bm{E}$ is the chromoelectric field $E^i=F^{i0}$, $D_0=\partial_0 +i g_s A_0$ and the matching coefficients of the chromoelectric sinlget-to-octet and octet-to-octet transitions are set to unity. The dots stand for higher order terms in the multipole expansion in the relative distance $r$.  

In a similar way, the pNREFT for the particle-particle sector is written in terms of antitriplet and sextet field 
\begin{equation}
    \psi(t,\bm{x}_1)  \psi(t,\bm{x}_2) \sim \sum_{\ell=1}^3 {\rm{T}}^\ell(\bm{r},\bm{R},t) \bm{{\rm{T}}}^\ell_{ij} +   \sum_{\ell=1}^6 \Sigma^\sigma(\bm{r},\bm{R},t) \bm{{\rm{\Sigma}}}^\sigma_{ij} \, ,
\end{equation}
with $\ell=1,2,3$, $\sigma=1,...,6$ and $i,j=1,2,3$ and the tensors $\bm{{\rm{T}}}^\ell_{ij}$ and $\bm{{\rm{\Sigma}}}^\sigma_{ij}$ can be found in ref.\cite{Brambilla:2005yk}. In the static limit the Lagrangian for two equal mass coloured fermions  reads~\cite{Brambilla:2005yk}
\begin{eqnarray}
\mathcal{L}^{F\bar{F}}_{\hbox{\tiny pNRQCD}}&=&\int d^3 r {\rm{Tr}} \left\lbrace {\rm{T}}^{\dagger} \left[ i D_0 - \mathcal{V}_{{\rm{T}}} -\delta M_T\right] {\rm{T}} + \Sigma^{\dagger} \left[ i D_0 - \mathcal{V}_\Sigma -\delta M_\Sigma \right] \Sigma \right\rbrace \nonumber \\
&+& \int d^3 r \sum_{a=1}^8 \sum_{\ell=1}^3 \sum_{\sigma=1}^6 \left[  \left( \bm{{\rm{\Sigma}}}^\sigma_{ij} T^a_{jk} \bm{{\rm{T}}}^\ell_{ki}  \right) \Sigma^{\sigma \dagger} \bm{r} \cdot g_s \bm{E}^a \, {\rm{T}}^\ell -  \left(\bm{{\rm{T}}}^\ell_{ij}  T^a_{jk}   \bm{{\rm{\Sigma}}}^\sigma_{ki} \right) {\rm{T}}^{\ell \dagger} \bm{r} \cdot g_s \bm{E}^a \, \Sigma^\sigma \right] 
\nonumber 
\\
&+& \cdots \, ,
\label{pNRQCD_FF}
\end{eqnarray}
where the dots stand for higher terms in the multiple expansion. 
In the high-temperature limit, the $r$-dependent potentials originating from a HTL resummed gluon are (we write them setting explicitly $N_c=3$)
\begin{eqnarray}
&& \mathcal{V}_{s}(r)=-\frac{4}{3}\alpha_s  \left[ \frac{e^{-m_D r}}{r} - i T \Phi(m_D \, r) \right] \, ,
\label{HTL_s}
\\
&& \mathcal{V}_{{\rm{T}}}(r)=- \frac{2}{3} \alpha_s  \left[ \frac{e^{-m_D r}}{r} - i T \Phi(m_D \, r) \right] \, ,
\label{HTL_t}
\\
&& \mathcal{V}_{o}(r)=\frac{1}{6} \alpha_s  \left[ \frac{e^{-m_D r}}{r} - i T \Phi(m_D \, r) \right] \, ,
\label{HTL_o}
\\
&& \mathcal{V}_{\Sigma}(r)=\frac{1}{3} \alpha_s  \left[ \frac{e^{-m_D r}}{r} - i T \Phi(m_D \, r) \right] \, ,
\label{HTL_Se}
\end{eqnarray}
where the function $\Phi(x)$ is given by\cite{Burnier:2007qm}
\begin{equation}
\Phi(x)=\frac{2}{x} \int_{0}^{\infty} dz \frac{ \sin(xz)}{(1+z^2)^2} \, ,
\end{equation}
whereas the $r$-independent mass and widths read the same for all the colour combinations\cite{Burnier:2007qm,Brambilla:2008cx}
\begin{equation}
\delta M_s=\delta M_o=\delta M_T=\delta M_\Sigma = -\frac{4}{3}\alpha_s (m_D+iT) \, .    
\end{equation}

In this work, we make a step further towards a relaxation of the high-temperature regime and we add the study of the thermal potential when the temperature scale is similar to the typical inverse size of the bound state. Moreover, as we have done in\cite{Biondini:2018ovz}, we include the contribution from gluo-dissociation in the singlet potential and we add the corresponding contribution for the antitriplet. In order to consider the latter process, we shall need the chromoelectric transitions in the second line of Eqs.~(\ref{pNRQCD_FbarF}) and (\ref{pNRQCD_FF}) respectively. 

In what follows, our focus will be on the attractive potentials. Indeed, a suppressed population of heavy pairs in the repulsive configurations is expected with respect to the pairs in a bound state, in the regime where the temperature goes down to $T \siml M_F v^2$. One may see this by looking at the Boltzmann distribution of the heavy pair, that for $M_F \gg T$ is well approximated by $e^{-(2M_F \pm |E'|)/T}$, where the plus sign stands for the octet and the minus sign for the singlet and $|E'| \approx M_Fv^2$. Hence, octet states are suppressed at small temperatures and their impact on the annihilation dynamics is expected to be marginal. The same argument applies for the relative importance of antitriplet over sextet states. In addition, former studies showed that the repulsive potentials have a moderate impact on the corresponding thermally averaged Sommerfeld factors, namely they remain close to unity for a rather broad range of temperatures~\cite{Biondini:2017ufr,Biondini:2018pwp}, at least for the HTL potentials. Nevertheless, a more careful analysis of this aspect is desirable in order to achieve a fully reliable and complete theoretical description, and we leave it for future work. Here, we use the results in the high-temperature limit for the octet and sextet potentials in Eqs.~(\ref{HTL_o}) and (\ref{HTL_Se}) in order to extract the corresponding spectral functions.

\subsubsection{Real part of the potential for $M \alpha_s \approx \pi T$}
\label{sec:real_part_gluo}
Thanks to Eqs.~(\ref{SH_like_1}) and (\ref{SH_like}) we account for non-perturbative effects like the Sommerfeld enhancement and bound-state formation at different temperatures. In practice we resum an infinite series of ladder diagrams where the exchanged gluon comes with a thermal propagator. In previous works\cite{Kim:2016kxt,Kim:2016zyy,Biondini:2017ufr,Biondini:2018pwp,Biondini:2018xor} the potentials and widths, often referred to as real and imaginary part of a thermal potential respectively, correspond to the high-temperature limit (i.e.~a HTL resummed gluon propagator). It accounts for a Debye-screened potential and the Landau damping, i.e.~soft $2 \to 2$ scatterings with light particles in the medium.
However, the temperature is evolving with time and we probe different distance scales by solving the Schr\"odinger equation.  We aim at relaxing this approximation by using a thermal potential and a width valid for $\pi T \approx M_F \alpha_s$. More specifically, we consider the hierarchy of scales $M_F \gg M_F \alpha_s \sim \pi T \gg m_{D} \gg \Delta V$. Here $\Delta V$ is understood as the difference between the static potentials.  We shall use this new result to better estimate the contribution from the attractive potentials, as far as the contribution from a resummed gluon propagator is concerned.  The imaginary part has been derived in ref.\cite{Brambilla:2013dpa}, whereas the real part has not been previously carried out. The latter has been computed for the abelian case, as given in ref.\cite{Escobedo:2008sy}, and  we can use it as a reference for the fermionic contribution to the gluon thermal self-energy. 

We give some details of the derivation in the Appx~\ref{App_pNRQCD_1} together with details on the running of the strong coupling $\alpha_s$. Here, we simply give the result. The temperature and the typical inverse size of the heavy pair are integrated out at the same time. One obtains a leading in-vacuum Coulomb potential with the following thermal contributions. We write the fermionic and bosonic contributions to the potential separately. The former is found to be 
\begin{eqnarray}
&&\delta V^{q}(r,T) = -\frac{C_F}{4} \, \alpha_s  \, r \, m^{2}_{D,q}   - C_F \frac{3}{2\pi} \, \alpha_s \, r^2 T \, m^{2}_{D,q} \,\zeta(3)  + C_F \frac{ \alpha_s \, m^{2}_{D,q}}{4 \,  \pi^2 r T^2} \int^{\infty}_{0} \frac{dx \, F^q(xrT) }{x\left( e^{x/2} + 1 \right) } \, ,
\label{result1_mpi}
\nonumber \\
\phantom{s}
\\
&&F^q(u)=\left[ -4 -3u^2 + (u^2 + 4 ) \cos(u)  +   u \sin(u) + ( 6 u + u^3 )\, {\rm{Si}}(u) \right] ,
\label{int_q_mpi}
\end{eqnarray}
where the superscript stands for the quark contribution to the gluon self-energy and $m^{2}_{D,q} = (g_s^2 T^2 N_f T_F )/3$, $\zeta$ is the Rienmann zeta function and Si$(u)=\int_0^u \sin(t) /t $ is the sine-integral function.  
The result agrees with the abelian version in ref.\cite{Escobedo:2008sy}.\footnote{After correcting a typo in Eq.~(16) of the paper.}
The bosonic contribution to the thermal gluon self-energy reads
\begin{eqnarray}
&& \delta V^{g}(r,T) = -\frac{C_F}{4} \alpha_s  \, r \, m^{2}_{D,g}  - C_F\frac{\alpha_s \, r^2 T \, m^{2}_{D,g}}{ \pi} \, \zeta(3)  + C_F \frac{ \alpha_s \, m^{2}_{D,g}}{8 \,  \pi^2 r T^2} \int^{\infty}_{0} \frac{dx \, F^g(x \, rT)}{x\left( e^{x/2} - 1  \right)}  \, ,
\label{result2_mpi}
\nonumber
\\
\phantom{x}
\\
&& F^g(u)=\left[ -22 -3u^2 + (u^2 +10 ) \cos(u) + \left( u + \frac{12}{u} \right)  \sin u +  (u^3 +12 u)  {\rm{Si}}(u) \right] \, .
\label{int_g_mpi}
\end{eqnarray}
where the superscript stands for the bosonic contribution to the gluon self-energy and $m^{2}_{D,g} = (g_s^2 T^2 N_c)/3$.

 In order to check our results, we considered the limits $r\, \pi T \ll 1$ and  $r\, \pi T \gg 1$ and compare with the literature\cite{Brambilla:2008cx}.  First, we perform the limit  $r \, \pi T \ll 1$ that amounts to expanding the functions $F^q$ and $F^g$ in Eqs.~(\ref{result1_mpi}) and (\ref{result2_mpi}) for small values of their arguments. Then we sum $\delta V^{q}(r,T)$ and $\delta V^{g}(r,T)$ and one recovers the known potential as in \cite{Brambilla:2008cx}. 
Second, the complementary situation $\pi T \gg 1/r \gg m_D$ is recovered with a numerical check. Indeed, the analytical limit is not straightforward and one should go back to the original loop integral and implement an integration by regions. We find easier to compare numerically the limit $r \pi T \gg 1$ of Eqs.~(\ref{result1_mpi}) and (\ref{result2_mpi}) with the corresponding result given in ref.\cite{Brambilla:2008cx}.
We remark that the high-temperature singlet potential, i.e.~real part of (\ref{HTL_s}), is only recovered from our results here when $1/r \gg m_D$ is implemented in the same equation (\ref{HTL_s}). Relaxing more than one hierarchy at a time is challenging and beyond the scope of the present work. The potentials (\ref{result1_mpi}) and (\ref{result2_mpi}) are also valid for the antitriplet case when accounting for the change $C_F \to C_F/2$. 

\subsubsection{Antitriplet to sextet transitions at finite temperature}
\label{sec:gluodissociation}
When DM is a real scalar, particle-particle (antiparticle-antiparticle) $t$-channel annihilations processes into light degrees of freedom are possible for the coannihilation partner, namely $FF \to qq $ ($\bar{F}\bar{F} \to \bar{q}\bar{q} $). A similar situation occurs in the simplified model with Majorana fermion DM and an accompanying coloured scalar\cite{Garny:2014waa,Garny:2015wea}. Accordingly, an attractive antitriplet and a repulsive sextet potentials can be relevant for determining the annihilation cross section. Besides the effect of Yukawa screened potentials and Landau damping as described in Sec.~\ref{sec:thermal_potentials_NREFT}, an additional modification to the thermal potentials can occur. Making again the analogy with heavy quarkonium at finite temperature, we compute here the analogue of the singlet-to-octet thermal break up~\cite{Brambilla:2008cx, Brambilla:2011sg}, which is also known in the literature as gluo-dissociation\cite{Kharzeev:1994pz}. The bound state can be broken by a collision with a thermal gluon from the bath, that carries energy/momenta of order of the binding energy $M_F \alpha_s^2$, and induces the transition to an unbound colour octet state. 
In the model considered here, the transition between antitriplets and sextets mediated by a thermal gluon is relevant as well, which shares many similarities with the singlet-to-octet thermal process. To the best of our knowledge, antitriplet-to-sextet transitions at finite temperature have not been addressed. 

For the sake of the numerical extraction of the spectral functions in (\ref{SH_like_1}), we consider the antitriplet-to-sextet transitions in the static limit. This amounts to taking the $M_F \to \infty$ limit. It has been pointed out that the relativistic corrections are of the same order of the static contribution for the thermal widths (or the imaginary part of the thermal potential) in ref.\cite{Brambilla:2011sg}. However, such derivation fully exploit a calculation of expectation values for the singlets in the pNRQCD, and it is not clear how to use such a result in the setting of a plasma-modified Schr\"odinger equation as in Eq.~(\ref{SH_like_1}).\footnote{An interesting approach for connecting expectation values in pNRQCD and bound-state formation/dissociation involves an open-quantum-system approach\cite{Brambilla:2016wgg,Brambilla:2017zei}.} Therefore, we stick to the gluo-dissociation in the static limit, extract a potential as a matching coefficient of the pNRQCD and include it in our extraction of the Sommerfeld factors. We leave the derivation of the antitriplet-to-sextet transition beyond the static limit for future work.

\begin{figure}[t!]
\centering
\includegraphics[scale=0.55]{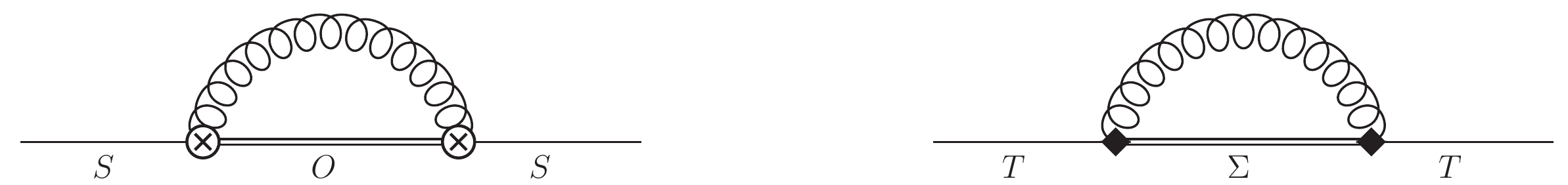}
\caption{\label{fig:SO_TS}The single continuous line stands for a singlet (antitriplet) propagator, the double line for an octet (sextet) propagator, the curly line for a chromoelectric correlator and the circle with a cross for a chromoelectric dipole vertex between singlet and octet (antitriplet and sextet) fields.}
\end{figure}

The calculation closely follows the one for the singlet-to-octet transitions in ref.~\cite{Brambilla:2008cx,Brambilla:2011sg}, that amounts to computing the one-loop self-energy for the singlet field (Fig.~\ref{fig:SO_TS}, diagram on the left). Similarly, we compute the one-loop graph in Fig.~\ref{fig:SO_TS} (right diagram) where a chromoelectric thermal gluon induces the transition from an antitriplet to a sextet field. In this setting, the hierarchy of scale reads $M_F \gg M_F\alpha_s \gg \pi T \approx \Delta V' \gg m_D$. Then, we start with the pNRQCD for the $FF$ sector in Eq.~(\ref{pNRQCD_FF}), compute the one-loop diagram and obtain the correction to the in-vacuum potential as follows 
\begin{eqnarray}
\delta \mathcal{V}_{\rm{T}} &=& -i g_s^2 N_c r_i \int \frac{d^4 k}{(2\pi)^4}  \frac{i}{-k^0-\Delta V' + i\epsilon} \, \left[ (k^0)^2 \, D^{(0)}_{ij} (k) + k_i k_j \, D^{(0)}_{00}(k) \right] \, r_j   \, , \nonumber 
\\
&=& \frac{4N_c}{3\pi} r^2 T^2 \Delta V' f(\Delta V'/T) - i \frac{2 N_c}{3} \alpha_s r^2 (\Delta V')^3 n_B(\Delta V') \, ,
\label{triplet_gluo_diss}
\end{eqnarray}
where $D^{(0)}_{00}(k), D^{(0)}_{ij}(k)$ is the unresummed gluon propagator in the Coulomb gauge~\cite{Landshoff:1992ne}, $\Delta V'=H_\Sigma-H_{\rm{T}} \approx V_\Sigma - V_{\rm{T}} \approx \alpha_s/r$ being the difference of the antitriplet and sextet energies in the static limit\footnote{We label the sextet-antitriplet energy difference with a prime to distinguish it from octet-singlet energy difference in the literature\cite{Brambilla:2008cx}.},  $n_B(x)=1/(e^{x/T}-1)$ is the Bose-Einstein distribution and
\begin{equation}
    f(t)=\int_{0}^{\infty} dx \frac{x^3}{e^x-1} {{\rm}P} \, , \frac{1}{x^2-t^2} \, ,
\end{equation}
with P standing for the principal value. When the temperature is larger than the difference of the static energy, the thermal potential reads
\begin{equation}
    \delta \mathcal{V}_{{\rm{T}}}=\frac{2N_c}{9} \alpha_s^2 r T^2-i \frac{2 N_c}{3} \alpha_s^3 T \, .
\end{equation}

\subsubsection{Numerical implementation of the thermal potentials}
\label{sec:num_potential}

Before we use the potentials for the different colour configurations to extract the spectral functions from eqs.~(\ref{SH_like_1}) and (\ref{SH_like}) some practical remarks are in order.

For the repulsive potentials we use the HTL expressions in eq.~(\ref{HTL_o}) and (\ref{HTL_Se}) for any temperature. 
It is not strictly rigorous to use the HTL potentials at small temperatures, namely $M_F\alpha_s \gg \pi T$, 
but we believe this is a  reasonable approximation. The solution of Eq.~(\ref{SH_like_1}) is well-behaved at all distances and the deviation from the in-vacuum potentials decreases with temperature thus limiting the impact in the regime where its use is most questionable.\footnote{In the regime of very small temperatures, 
$\pi T \siml M \alpha_s^2$, we multiply thermal interaction rates by the Boltzmann factor $\theta(-E')e^{-|E'|/T}$, in order to account for the exponential suppression below threshold in that temperature regime\cite{Biondini:2017ufr, Biondini:2018pwp}.}
As the contributions that get corrected by the repulsive potentials only account for a small part of the overall cross section any mismodelling of this part is naturally suppressed and  has only a minor impact of the relic density.   

The treatment of the attractive potential, i.e.~the single and antitriplet configuration, requires more care. We include Debye screening and Landau damping, and bound-state dissociation/formation via gluon radiation. These are two separate processes that have typically complementary regimes of relevance \cite{Burnier:2007qm,Brambilla:2008cx,Brambilla:2010vq,Brambilla:2011sg}. As the freeze-out scans a broad range of temperatures, the ordering of the relevant scales may change. This ought to be reflected in the potentials used for the derivations of the spectral functions.
The typical in-vacuum energy scale relevant in the Schr\"odinger equation is set by the inverse size of the pair. For a Coulomb-like state this corresponds to the inverse Bohr radius $1/a_0 = M_F \alpha_s C_{{\rm{F}}}/2 \approx M_F \alpha_s$. 
We obtain an estimate for the Bohr radius at different mass scales $M_F$ by using a recursive relation $1/a_0 = M_F C_{{\rm{F}}}/2 \alpha_s(1/a_0)$ with a running $\alpha_s$ as given in the Appx~\ref{App_pNRQCD_1}.
This scale can now be compared with the largest thermal scale in our setting, $\pi T$ (by assumption $m_D \sim g_sT < \pi T$ in a weak coupling framework) and we select the potentials based on the relation between them. 
For Debye screening and Landau damping, we use the HTL expressions (\ref{HTL_s}) and (\ref{HTL_t}) when $\pi T >M_F \alpha_s$ and switch to the sum of eq.~(\ref{result1_mpi}) and (\ref{result2_mpi}) when $\pi T \leq M_F \alpha_s$. Note that the in-vacuum Coulomb potentials for the singlet and antitriplet have to be added in the latter case because eq.~(\ref{result1_mpi}) and (\ref{result2_mpi}) represent a correction to it. 
For gluodissociation, we include the corresponding potentials when $\pi T \leq M_F \alpha_s$ is satisfied.  This is necessary to justify the use of the potential in the first place. For the antitriplet, the real and imaginary part of the potential is given in eq.~(\ref{triplet_gluo_diss}), whereas for the results for the singlet can be found in ref.\cite{Brambilla:2008cx}.

Of course, this is an approximation since not all scales are considered separately. In particular the arrangement of the smaller scales $\Delta V, \Delta V' \approx M_F \alpha_s^2$ and $m_D$ is not fixed and may differ from the choice we made in our computation. A more systematic treatment of the potentials is desirable and could be achieved by following the path outlined in the heavy quarkonium literature. However, it is not only a more complete and consistent set of potentials that have to be derived, but rather a more appropriate way to connect computations in the pNRQCD with rate equations. For example, when $\Delta V \gg m_D$, the extraction of a potential to be plugged into a Schr\"odinger equation is not strictly rigorous. One should rather compute the energy and width of a bound state in  pNRQCD directly (see e.g.~\cite{Brambilla:2010vq,Brambilla:2013dpa}) which changes the whole procedure for the derivation of Sommerfeld factors at finite temperature. We leave these considerations for future work. Here, we pursued a more phenomenological recipe and use different potentials at finite temperature to address the impact of the attractive ones on the relic density for the model at hand.

\subsection{$S \to F$ conversion rate}
\label{sec:CONV_rate}
An essential assumption of the coannihilation mechanism is chemical equilibrium between the DM and the coannihilation partner. If the two species are not in equilibrium the $S$ abundance cannot be depleted efficiently through $F \bar{F}$ annihilations and the freeze-out dynamics change considerably \cite{Garny:2017rxs}. In order to estimate whether the thermal contact between $S$ and $F$ is sufficient to couple the two reservoirs one has to consider the rate $\Gamma$ at which $S$ particles is converted into the mediators. As long as $\Gamma \geq H$ throughout the freeze-out it is reasonable to assume that  the $F$ particle does not evolve independently of the actual DM and our formalism applies.

\begin{figure}[ht!]
    \centering
    \includegraphics[scale=0.7]{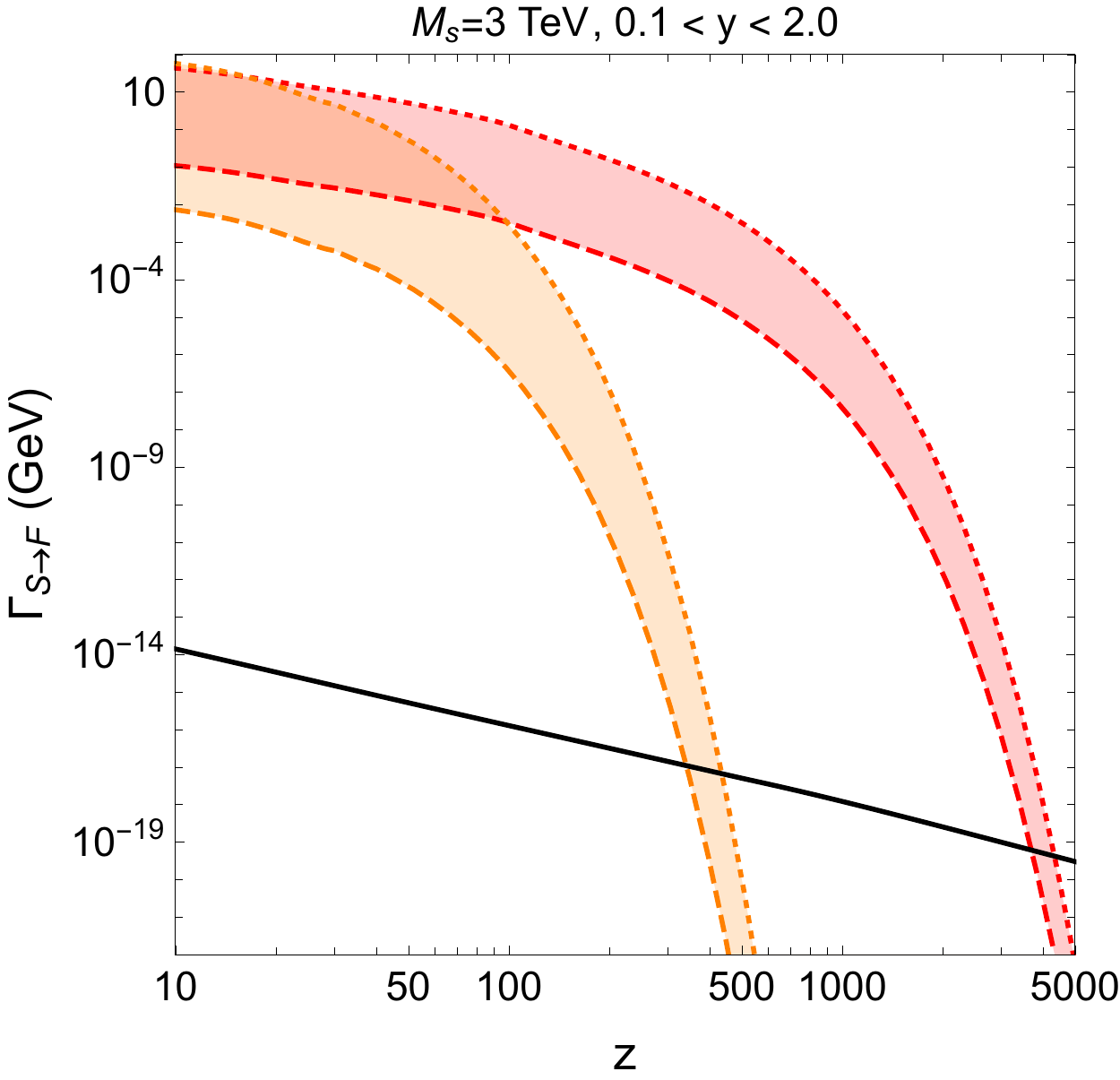}
    \caption{The sum of the $2 \to 1$ and $2 \to 2$ rates compared with the Hubble rate for $M_S=3$~TeV, $\Delta M/M_S=0.1$ (orange) and $\Delta M/M_S=0.01$ (red), and for $0.1 \leq y \leq 2.0$.
    }
    \label{fig:conversion}
\end{figure}

Two processes contribute to the conversion of a DM scalar particle into a coloured fermion and vice versa. There are $2 \to 1$ and  $2 \to 2$ processes: the former is the inverse decay of the DM and SM quark into $F$ ($S q \to F$) whereas the second is a scattering process with light partons in the medium ($S q(g) \to F g (q)$). The two rates can be interpreted as the imaginary part of the real scalar self-energy where the SM quark propagator is treated at leading order or resummed respectively. 
The result for the $2 \to 1$ thermal interaction rate reads:
\vspace{0.2 cm}
\begin{equation}
\begin{minipage}[c]{0.2 \linewidth} \includegraphics[scale=0.6]{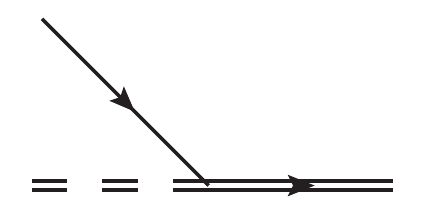} \end{minipage} \phantom{sssss} \Gamma_{2 \to 1} = \frac{|y|^2 N_c M_S}{4 \pi} \left( \frac{\Delta M}{M_S} \right)^2 n_F(\Delta M) \, ,
\label{2to1}
\end{equation}
where $n_F(x)$ is the Fermi-Dirac distribution $n_F(x)=1/(e^{x/T}+1)$. 
This rate vanishes for $\Delta M \to 0$ and it is exponentially suppressed in the regime $\Delta M \gg T$. 
The light quark from the medium can interact with $S$ and add enough energy to produce a slightly heavier $F$ particle. At temperatures smaller than $\Delta M$, the light quarks from the bath have typical energies smaller than the mass splitting, inducing the Boltzmann suppressed behaviour of the process. The rate for the second process read
\begin{equation}
\begin{minipage}[c]{0.2 \linewidth} \includegraphics[scale=0.6]{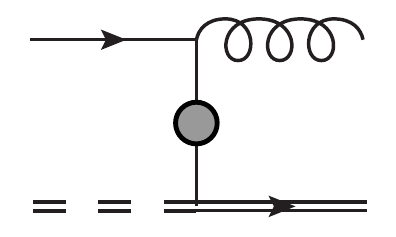} \end{minipage} \phantom{sssss}
\Gamma_{2 \to 2}= \frac{|y|^2 N_c }{8 M_S} \int_{\bm{p}} \frac{\pi m_q^2 n_F\left(\Delta M + \frac{p^2}{2 M_S}\right)}{p(p^2+m_q^2)} \, ,
\label{2to2}
\end{equation}
where $p=|\bm{p}|$ and the thermal quark mass $m_q=g_s^2 T^2 C_F/4$. This rate is the dominant one at large temperatures and we consider HTL resummation to obtain the correct result, especially in the case of light quarks. On the other hand, the $2 \to 1$ process is more relevant at small temperatures.
We stress that the in-vacuum mass has been set to zero to simplify the calculation of the rates, and this is a good approximation as long as $m_q^{T=0} \leq \pi T$.

In Fig.~\ref{fig:conversion}, we show the sum of the two rates in Eq.~(\ref{2to1}) and (\ref{2to2}) compared to Hubble rate for $M_S=3$ TeV, for two relative mass splittings $\Delta M/M_S=0.1$  and $\Delta M/M_S=0.01$, and for $y \in [0.1,2.0]$ as considered in this work. This leads to the orange and red bands in Fig.~\ref{fig:conversion}. One can see that smaller mass splittings allow for maintaining the chemical equilibrium in the dark sector down to smaller temperatures.

\section{Phenomenology}
\label{sec:Pheno}

In this section we will discuss the phenomenology of the model and introduce the most relevant experimental constraints. Since the mediator is a colour charged fermion it has a substantial production cross section at the LHC. Once produced, the mediator will either decay to DM and light quarks or to DM in association with top quarks, depending on the coupling structure. In addition, the DM can also interact with quarks at much lower energy scales and thus we expect a signal at direct detection experiments. Indirect searches for DM annihilations happening in the Universe today are also a possible way to test thermal DM. However, the coannihilation mechanism discussed here generically predicts an annihilation cross section smaller than the canonical value of $\sigma v \approx 3 \times 10^{-26} \mbox{cm}^3/s$. As indirect detection generally struggles to probe $\sigma v \approx 3 \times 10^{-26} \mbox{cm}^3/s$ in the mass range of interest here, i.e. $M_S\geq 500$ GeV, we do not discuss it further. For a more detailed discussion of indirect detection in coannihilation scenarios see for example \cite{Biondini:2018ovz}.

\subsection{LHC searches}

\begin{figure}[tb]
    \centering
    \includegraphics[width=0.8\textwidth]{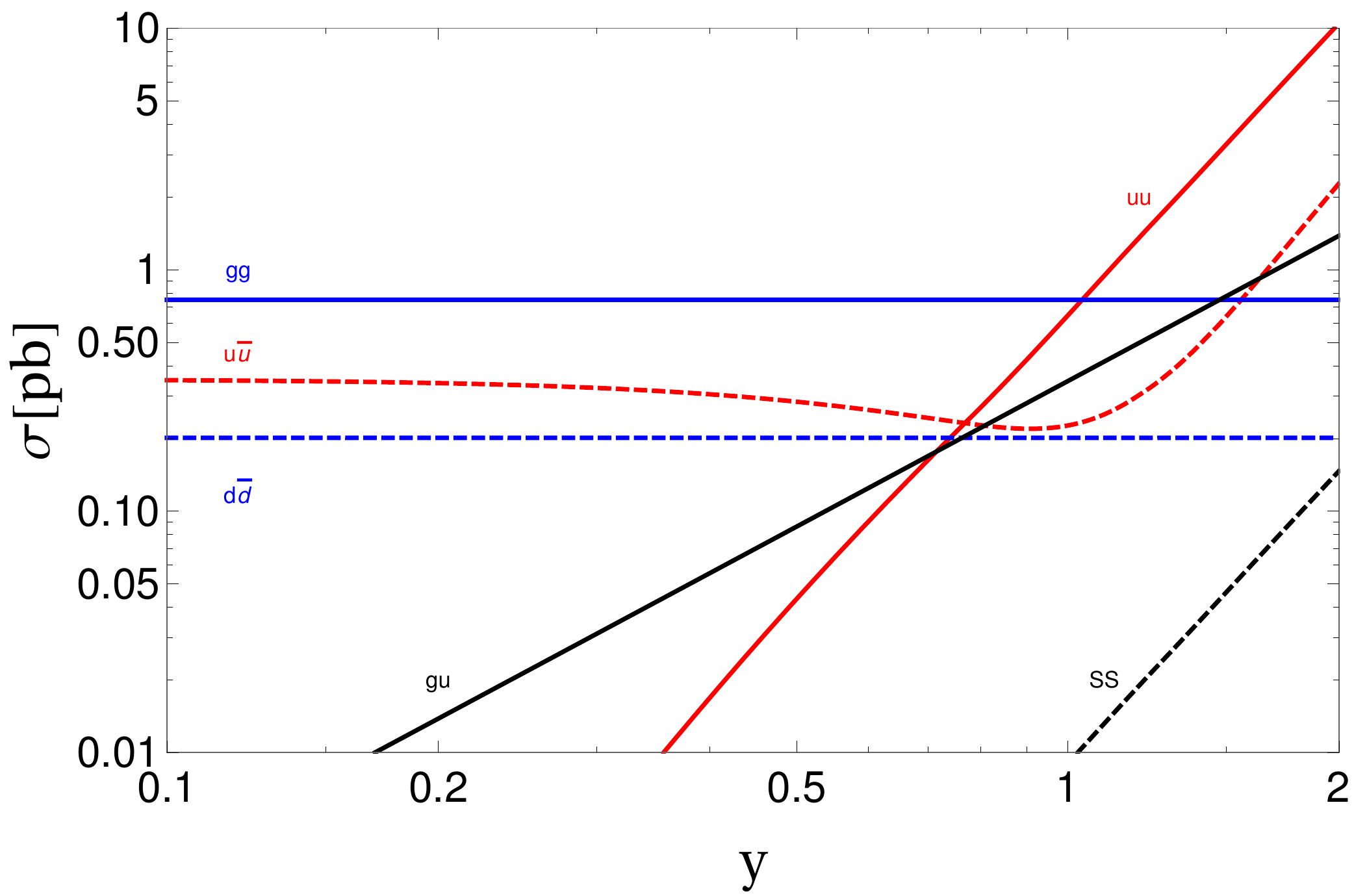}
    \caption{New physics production cross sections at the $\sqrt{s}=13$ TeV LHC as a function of the Yukawa coupling $y$ for a representative choice of the masses ($M_S=500$ GeV and $M_F=550$ GeV). The pure gauge mediated production of mediator pairs from $gg$  ($d\bar{d}$) initial states is shown by the blue solid (dashed) line. The mixed gauge and new physics mediated production from $u\bar{u}$ is indicated by the dashed red line while the exclusively new physics induced cross section for particle-particle production from $uu$ initial states is marked by the solid red line. The cross section of mixed DM mediator final states from $gu$  is shown as black, solid. For comparison we also show the direct $SS$ production cross section as black, dashed.  \label{fig:LHC_xsec}}
\end{figure}

LHC searches are an attractive way to search for DM with colour charged mediators \cite{An:2013xka,Chang:2013oia,Garny:2014waa,Giacchino:2015hvk}. In general there are various ways to produce DM and the mediator at colliders. The relative importance of the different production modes depends on the parameters of the theory and the flavour of the quark the  DM couples to. In the following, we will first discuss the case of light quarks before turning to DM coupling to tops. 

\subsubsection{Valence quarks}

In this case, both gauge interactions and the new physics coupling $y$ can dominate the production cross section of new physics.
In particular the dependence on $y$ is very strong for DM coupling to light quarks, see Fig.~\ref{fig:LHC_xsec} for an illustration for a representative benchmark point. 
As expected, the biggest cross section for small to moderate $y$ comes from the pure QCD production of the coloured mediators from gluons and $q\bar{q}$ initial states. At $M_F=550$~GeV this cross section is  $\approx 1$ pb which allows for ample production at the LHC. The amplitude for $F\bar{F}$ production from the quark flavour the DM couples to also receives a contribution from DM exchange which interferes destructively with the QCD part at moderate $y$. At even larger $y$ the cross section of $F\bar{F}$ is dominated by the $t$-channel $S$ exchange. 
However, in this regime the production rate receives a big contribution from particle-particle production which is also mediated by the DM. At large $y\geq 2$ this contribution dominates the total new physics production rate. This behaviour can be understood by considering the parton distribution functions (PDFs) at the relevant centre-of-mass energies. In $\sqrt{s}=13$ TeV collisions, the PDF of the up-quark typically exceeds the PDF of $\bar{u}$ significantly and enhances the parton luminosity of the $uu$-initial state relative to $u\bar{u}$.
In addition, it is also possible to produce the DM directly, either in pairs or in association with the mediator. The associate production rate is quite interesting as it is of comparable size as the leading mediator pair production cross sections at intermediate $y$. However, since this process scales as $\alpha_s y^2$  while  $FF$ production (and $F\bar{F}$ production in the large $y$ limit) scales as $y^4$ the relative importance of this channel declines at large $y$. DM pair production via $t$-channel $F$ exchange generally has the smallest cross section over the full range of $y$-values considered here. In addition, DM-pairs can also be produced by an off-shell Higgs due to the portal coupling $\lambda_3$. For reasonable values of $\lambda_3$ this cross section is substantially smaller than all other production modes discussed above \cite{Hessler:2014ssa} and we neglect it.

Naturally, the production cross section alone is insufficient to assess the observably of the signal and the final state needs to be considered in more detail. In general, the most promising collider signal in this model will consist of DM, i.e.~missing transverse energy ($MET$), in association with jets. The visible part of the final state can either come from the decay of the coannihilation partner $F$ or from additional radiation emitted by the initial or final state.  The LHC has excellent sensitivity to multijet+$MET$ final states provided that the events are energetic enough. In the mono-(multijet-)search at ATLAS the minimal requirement is one (two) jet(s) with a transverse momentum $p_T\geq 250$ GeV~\cite{Aaboud:2017phn,Aaboud:2017vwy}.  In the case at hand, such energetic jets are not readily available since the mass splitting $\Delta M$ is substantially smaller than $250$~GeV in coannihilation scenarios. As the jets from the decay are comparatively soft, the efficiency of the multijet search suffers. Only boosted particles or events with additional hard radiation provide a signal above the analysis threshold. Therefore, the visible cross section is substantially smaller than the naive estimates shown above and the experimental efficiency exhibits a non-trivial dependence on the parameters of the theory.
Since the dominant production mode in our scenario can differ substantially from the one used in the experimental analysis we perform a full recast of the relevant experimental searches. We simulate the production of the mediator in association with up to two additional jets with Madgraph5@NLO~\cite{Alwall:2011uj} and hand the events to Pythia8~\cite{Sjostrand:2014zea} for hadronisation and fragmentation. The effects of the detector are approximated with Delphes3~\cite{deFavereau:2013fsa}. Finally, we use the Checkmate2~\cite{Dercks:2016npn} implementation of the ATLAS searches for monojet~\cite{ATLAS-CONF-2017-060,Aaboud:2017phn} and multijet~\cite{Aaboud:2017vwy} events with missing transverse energy to derive our limits.   
It turns out that the multijet search has the best sensitivity in the full parameter space under consideration here. 

\subsubsection{Top quark}

In the case of top quarks the situation is both simpler and more complicated than for DM coupling to valence quarks. On the one hand,  the top PDF is negligible at the energies reached by the LHC.  Therefore, only the gauge interactions are relevant  and particle-particle final states do not contribute to the cross section. On the other hand, in the region of parameter space relevant for coannihilation, the large mass of the top precludes two-body decays and we need to consider more complicate decays  such a $F \to SWb$ or even four-body decays $F \rightarrow Sb f f'$ where $f$ and $f'$ denote light SM fermions. 
Phenomenologically, this process is  similar to the stop-coannihilation region in simplified models of Supersymmetry~(SUSY)~\cite{Aaboud:2017aeu,Sirunyan:2018lul} and there exist limits on the pair production cross section of stop pairs $\Tilde{t} \Tilde{t}^\dagger$. We use the limits on stops-production reported by \cite{Aaboud:2017aeu} to estimate the LHC limits on the model under consideration here. We would like to caution that simplified SUSY is similar but  not identical to the fermionic $t$-channel mediator model. For instance, there are differences between the differential cross sections for the production of scalars and fermions. 
In addition, there is an angular correlation between the decay products of a pair of fermions which is absent in the case of decaying scalars.  Nevertheless, in light of the uncertainties introduced by a phenomenologist's modelling of detector effects, we do not see a clear advantage in recasting the full analysis. Unfortunately, the experimental sensitivity turns out to be too weak to constrain $M_F\geq 500$ GeV in region with small $\Delta M$ and we cannot exclude the parameter space under consideration here. 
 
\subsection{Direct detection}

Direct detection experiments aim to observe the scattering of DM on regular matter by searching for the recoil of SM particles in a low background environment. The scattering rate of DM on nuclei is tightly constrained and the best current limits reach scattering cross sections of $\approx 4\times 10^{-47} \;\mbox{cm}^2$ \cite{Aprile:2017iyp}. The limits on DM in the mass range considered here are somewhat weaker than the above value, mainly due to the lower DM number density. Nevertheless, direct detection experiments have sufficient sensitivity to probe scenarios with a suppressed scattering cross section.

\begin{figure}[tb]
\centering
\includegraphics[width=0.26\textwidth]{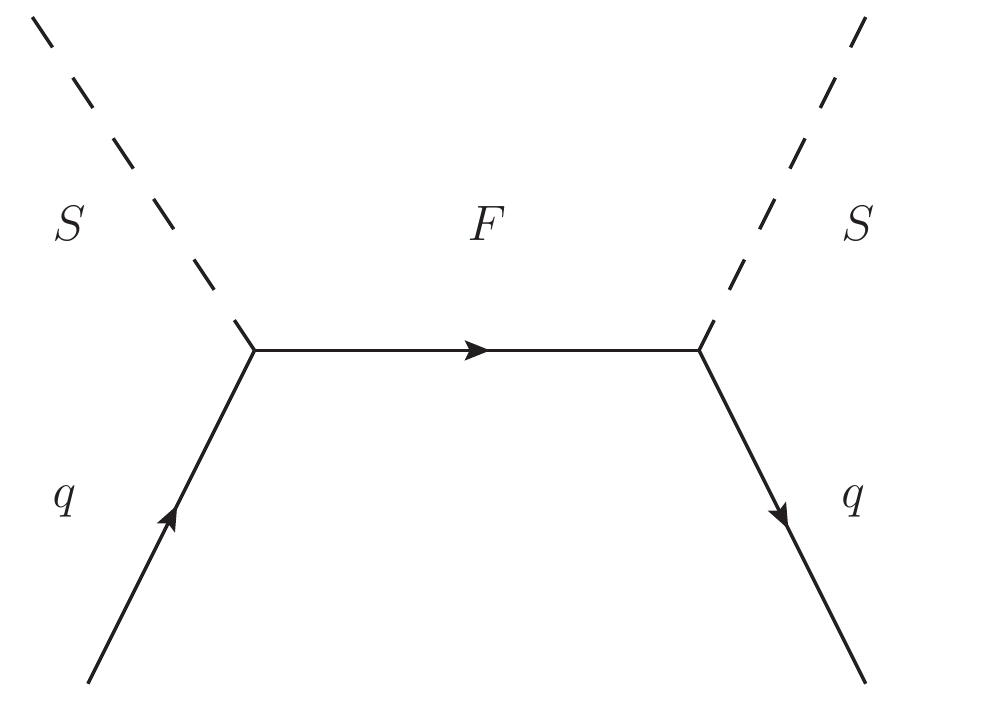}
\hspace{1.cm}
\includegraphics[width=0.22\textwidth]{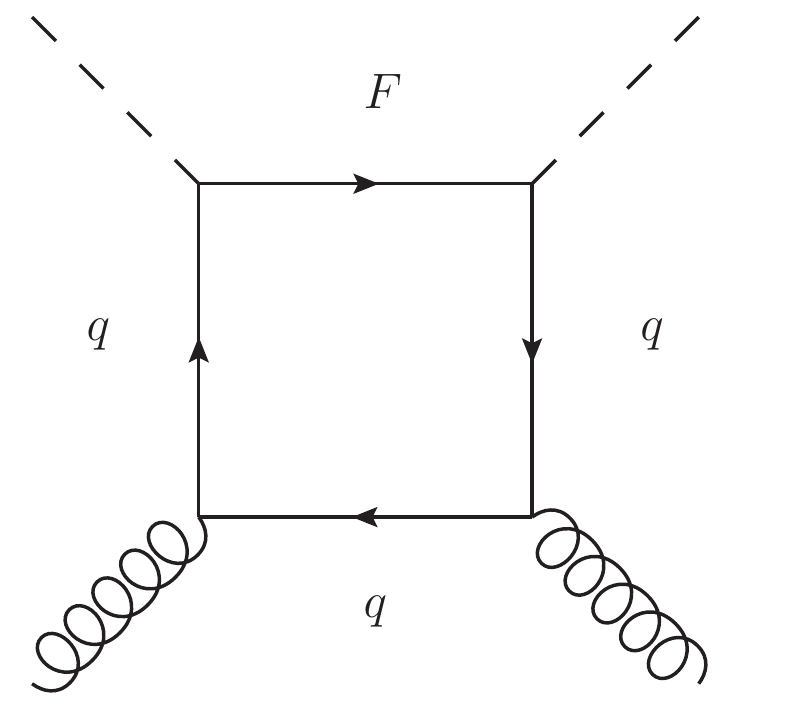}
\hspace{1.cm}
\includegraphics[width=0.25\textwidth]{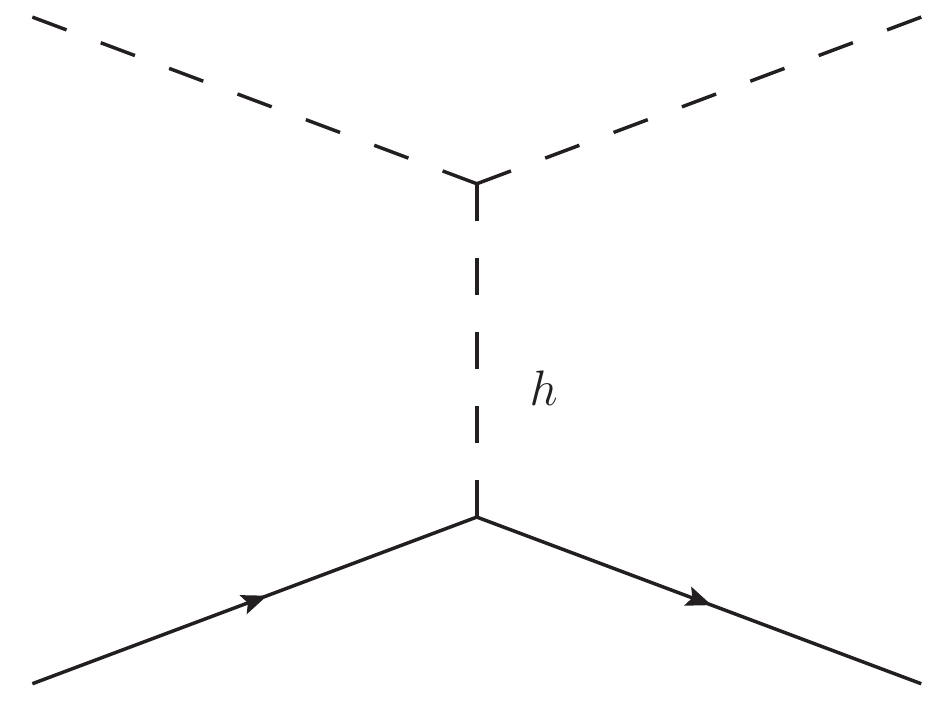}
\caption{Set of representative Feynman diagrams contributing to the effective DM nucleon interaction.\label{fig:DD}}
\end{figure}

The direct detection cross section of scalar DM on a nucleon $N$ is given by \cite{Hisano:2015bma}
\begin{align}
    \sigma_N = \frac{1}{\pi} \frac{m_N^2}{(M_{S} + m_N)^2} f^2_N \;,
    \end{align}
where the dimension-full coupling $f_N$ is the coupling of the effective DM-nucleon interaction
\begin{align}
    \mathcal{L}_{N}= f_N S^2  \bar{N} N\,.
\end{align}
The effective coupling receives contributions from mediator exchange $f_{N,y}$ and from the direct coupling of the DM to the Higgs $f_{N,\lambda_3}$ that add coherently $f_N= f_{N,y} + f_{N,\lambda_3}$, see  Fig.~\ref{fig:DD} for a representative set of the diagrams.
It is important to keep in mind that the nuclear scale $\mu_N=m_p$, which is relevant for direct detection, is vastly different compared to the scale of DM  annihilations $\mu_{ann}=M_S$. Consequently, the running of the couplings between these scales could have an impact on the phenomenology.
There are two different regimes of the running which have to be treated separately. Starting from the scale of annihilations,  we first encounter the regime below the DM mass scale and above the weak scale. Here, $S$ and $F$ are already heavy while SM states such as $H$  and $t$ are still light. 
At the weak scale the Higgs and top need to be integrated out and the operators are matched to a theory that is valid between the weak and the nuclear scale. 
As long as only one-loop running is considered, the low-energy operators relevant for direct detection can be formulated in an RGE-invariant way \cite{Hisano:2015bma,Hisano:2017jmz}. Therefore, we only need to calculate the running above the weak scale while the other contributions are already taken into account once we match the operators.

In the following we fill first discuss the low-energy formalism for direct detection.
Depending on the flavour of the quark that the DM couples to, the leading processes contributing to direct detection are different. In order to simplify the discussion we will focus on two extreme cases, i.e.~coupling to valence-quarks and coupling to the top-quark, and present these separately. 
Finally, we will discuss the connection between the weak scale and the annihilation scale and present our results for the RGEs in this regime.
%%%%%%%%%%%%%%%%%%%%%%%%%%%%%%%%%%%%%%%%%
\begin{subsubsection}{Valence quarks}
%%%%%%%%%%%%%%%%%%%%%%%%%%%%%%%%
The mediator induced coupling arises from two classes of diagrams. First, for valence quark tree-level exchange of $F$ between $Sq$ bilinears is possible, see Fig.~\ref{fig:DD} (left). In addition, box-diagrams of the type shown in Fig.~\ref{fig:DD} (middle) with the mediator and the quarks in the loop can generate an effective interaction with gluons.
For DM coupling to a light quark $q\in\{u,d,s\}$ the effective coupling $f_{n,y}$ reads \cite{Hisano:2015bma}
\begin{align}
       \frac{f_{N,y}}{m_{N}}= \frac{y_q^2}{M_{S}^2}\left[ \frac{1}{4} \frac{2 \tilde{r}^2 -1}{(\tilde{r}^2-1)^2}  f_{T_q} + \frac{3}{4}  \frac{1}{(\tilde{r}^2-1)^2}  (q(2)+\bar{q}(2))-\frac{1}{27}  \frac{1}{\tilde{r}^2-1} f_{T_G}\right]\;,
       \label{eq:DD_fermion}
\end{align}
where $\tilde{r}= M_F/M_S$. 
The first two terms in the above equation come from the tree-level exchange of the mediator. The last term is due to the loop-induced effective interaction with gluons.
The  effects of the nucleon structure are encapsulated in the coefficients $f_{T_q}$, $f_{T_G}$, $q(2)$ and $\bar{q}(2)$. The first coefficient $f_{T_q}=\langle N | m_q q \bar{q | N \rangle}/m_N$ parametrises the expectation value of the scalar quark bi-linear in the nucleon,   while $q(2)$ and $\bar{q}(2)$ are the second moments of the parton distribution of the quark and the anti-quark respectively. Finally, $f_{T_G} = 1 -\sum_{q=u,d,s} f_{T_q}$ is the contribution of the gluonic matrix element.
In our analysis we use the default values of micromegas5.0 \cite{micromegasmanual} for  $f_{T,q}$ and $f_{T_G}$, while the values for $q(2)$ and $\bar{q}(2)$ are taken from \cite{Hisano:2015bma}. The relative importance of the different contributions only depends on the mass ratio $\tilde{r}$. In the coannihilation regime, i.e. for $\tilde{r}\leq 1.2$, the effective coupling is dominated by the second contribution to Eq.~(\ref{eq:DD_fermion}) and the other terms are just a minor correction.

Last but not least, there is a contribution from Higgs exchange which is given by \cite{Cline:2013gha}
\begin{align}
    \frac{f_{N,\lambda_3}}{m_N}=\frac{1}{2} \lambda_3 \frac{1}{m_h^2} f_{h} \, ,
    \label{eq:DD_Higgs}
\end{align}
where $f_{h}$ denotes the effective Higgs nucleon coupling. This parameter can be expressed in terms of $f_{T_q}$ and reads
\begin{align}
    f_{h}=\frac{2}{9}+\frac{7}{9}\sum_{q=u,d,s}f_{T_q}\;.
\end{align}
As can be seen, $f_{N,\lambda_3}$ and $f_{N,y}$ have a rather different dependence on the parameters of the theory. 
While $f_{N,\lambda_3}$ is completely insensitive to the new physics scales $M_{S}$ and $M_F$, $f_{N,y}$ depends strongly on the mass of the DM and its mass splitting with the mediator. Consequently, the relative importance of mediator and Higgs  exchange for direct detection depends on the choice of the parameters.
%%%%%%%%%%%%%%%%%%%%%%%%%%%%%%%%%%%%%%%%%%
\end{subsubsection}
%%%%%%%%%%%%%%%%%%%%%%%%%%%%%%%%%%%%%%%%%%
\begin{subsubsection}{Top quark}
%%%%%%%%%%%%%%%%%%%%%%%%%%%
The general structure of  the DM nucleon coupling remains similar to the previous case. 
However,  $f_{N,y}$ is markedly different from the case of DM interacting with valence quarks.
Since the top is very massive, it is essentially absent in the proton and the tree-level exchange of $F$ does not contribute to direct detection. Therefore, the only $y$-dependent contribution is due to the loop induced DM gluon coupling depicted in the second panel of Fig.~\ref{fig:DD}. In contrast to the gluon contribution in Eq.~(\ref{eq:DD_fermion}), the quark is now also a heavy degree of freedom and $m_t$ needs to be taken into account in the loop computation. Consequently, the expression gets modified and one finds~\cite{Hisano:2015bma}
\begin{align}
       \frac{f_{N,y}}{m_{N}}= -\frac{1}{9} y_t^2 f_{T_G}\sum_{i=a,b,c} f^+_{i}(M_S,M_F,m_t)\;,
       \label{eq:DD_fermion_top}
\end{align}
where $f^+_i$ are loop functions which are given in the Appx. of~\cite{Hisano:2015bma}.
In addition, the Higgs mediated contribution $f_{N,\lambda_3}$ also contributes to the direct detection. The Higgs portal is completely independent of the flavour structure of the mediator interactions and $f_{N,\lambda_3}$ takes the same form as in Eq.~(\ref{eq:DD_Higgs}). 

\subsubsection{Running of the couplings}

We are interested in the RGE running of the couplings $y$ and $\lambda_3$ between the weak scale and the DM mass scale. One might expect that only the contribution of SM fields to the renormalization is relevant since both $F$ and $S$ are heavy. However, in the coannihilation regime the mass splitting $\Delta M\ll M_S$  is still a dynamical scale at intermediate energies. Therefore, some diagrams with internal $S$ and $F$ propagators are less suppressed than a naive simple power counting might indicate.  
In order to treat this subtlety, it is again convenient to resort to the non-relativistic effective field theory methods we have already employed in our treatment of DM annihilations.
In the following we will briefly introduce our approach and summarise our main results. A more detailed discussion can be found in Appx.~\ref{app:RGE}. 

We implement a $1/M_S$ expansion of the Lagrangian given in Eq.~(\ref{Lag_RT}) such that momentum modes of order $M_S$ are integrated out and work in a theory that includes $\phi$, the non-relativistic component of $S$, as well as $\psi$ and $\bar{\chi}$, the corresponding non-relativistic components of $F$. For direct detection, the most important operators are $\frac{y}{\sqrt{M_S}}\phi \bar{\psi} P_R q$ and $\frac{\lambda_3}{M_S}\phi^\dagger \phi H^\dagger H$ \footnote{We would like to remind the reader that in our convention the non-relativistic scalar field $\phi$ has mass dimension $3/2$.}. Apart from the $\bar{\psi}\psi H^\dagger H$ operator and the corresponding one containing $\chi$, whose coefficient is zero in the full theory, these are the lowest dimensional operators including the low-energy fields. Therefore, we only include counterterms up to mass dimension-$5$ and perform the wave-function and vertex renormalization at one-loop. The SM RGEs remain unaffected since the corresponding operators are dimension-4.  
Using dimensional regularization with $D=4-\varepsilon$  in the $\overline{\text{MS}}$ scheme, and $t = \ln \bar{\mu}^2$, we find:
\begin{eqnarray}
&&\partial_t |y|^2 = \frac{|y|^2}{(4 \pi)^2} \Bigg\{ -4 \frac{\Delta M}{M_S} |y|^2 (1+  N_c) + |h|^2 -3 g_s^2 C_F \Bigg\} \, , 
\label{run_3}
\\
&&\partial_t \lambda_3=\frac{1}{(4 \pi)^2}  \Bigg\{ \left[ 6 \lambda_1 - \frac{9 g_w^2}{4} -4  N_c |y|^2 \frac{\Delta M}{M_S} + N_c |h|^2 \right] \lambda_3 
-2  N_c |y|^2 |h|^2   \Bigg\} \, 
\label{run_6}\,,
\end{eqnarray}
for the new couplings. Of particular interest to us is the last term in (\ref{run_6}). This contribution to the running arises from the vertex renormalization of $\phi^\dagger\phi H^\dagger H$ due to box diagrams with the vectorlike-fermion and  quarks in the loop. Since the dynamical scale entering the internal propagator is not $M_S$ but rather $\Delta M$, these diagrams (one given by $\psi$ and one by $\chi$ in the loop) contribute to the running and generate a term that depends on $y$ and $h$ rather than $\lambda_3$. Consequently,  $\lambda_3$ can only  be (close to) zero at multiple scales if $|y|^2 |h|^2 $ is negligible. This is realised to excellent approximation for DM coupling to up quarks. However, for DM coupling to the top this is clearly not the case and we always expect a substantial running of $\lambda_3$.  
%%%%%%%%%%%%%%%%%%%%%%%%%%
\end{subsubsection}
%%%%%%%%%%%%%%%%%%%%%%%%%%%%%

\subsection{Parameter space of thermal DM}
Now all the ingredients for a study of the cosmologically preferred  parameter space of our model are assembled. In the most general case the model has four free parameters, the two masses $M_{S}$ and $M_F$ and two couplings $y$ and $\lambda_3$, that are relevant for the phenomenology.
As the Planck satellite has measured the relic abundance with percent-level accuracy, we can use this constraint to remove one parameter from the theory and express it in terms of the remaining ones. We chose to fix $y$ as a function of the masses and $\lambda_3$.
To further simplify the discussion we will restrict ourselves to slices through the parameter space. Since we are mainly interested in the dynamics of the  coannihilation regime, which is more sensitive to the masses than to the Higgs portal, we consider two choices of $\lambda_3$. We take a look at the pure fermionic $t$-channel model
and set $\lambda_3=0$ before we consider $\lambda_3=1.5$. These two cases bracket the effect of the Higgs portal on the phenomenology of the model.

Before discussing the results in any detail, we would like to comment briefly on two limitations of our approach and how these impact the considered parameter region. First, our formalism is tailored towards the accurate description of the non-perturbative dynamics relevant in the coannihilation region. We would like to caution that our results are less precise than the standard approach in the regime with no coannihilation since we work in a $1/M_S$ expansion. Therefore, we only consider the region of parameter space with $\Delta M /M_S \leq 0.2$, for larger mass splitting we refer the reader to the literature~ \cite{Giacchino:2015hvk,Colucci:2018vxz}. In addition, we only consider $M_S \geq 500$~GeV. This ensures that the freeze-out via late-stage bound-state formation does not become sensitive to temperatures that are too low for the reliable use of our perturbative QCD potentials.
One way to explore $M_S < 500$ GeV is to determine the effective Sommerfeld factors on a lattice, see for example \cite{Kim:2019qix} for recent work in this direction.

\begin{figure}[tb]
\centering
\includegraphics[width=0.49\textwidth]{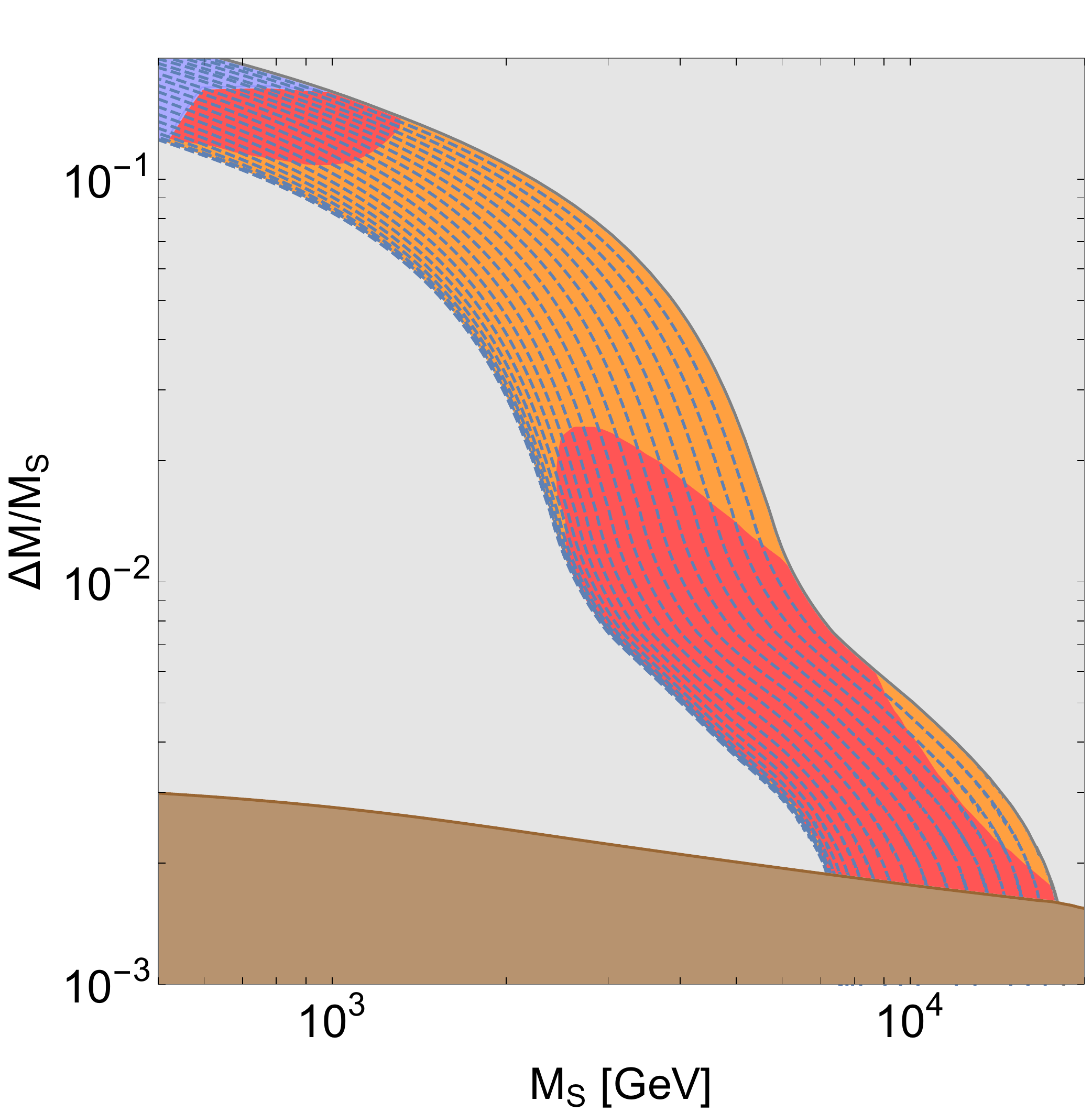}
\includegraphics[width=0.49\textwidth]{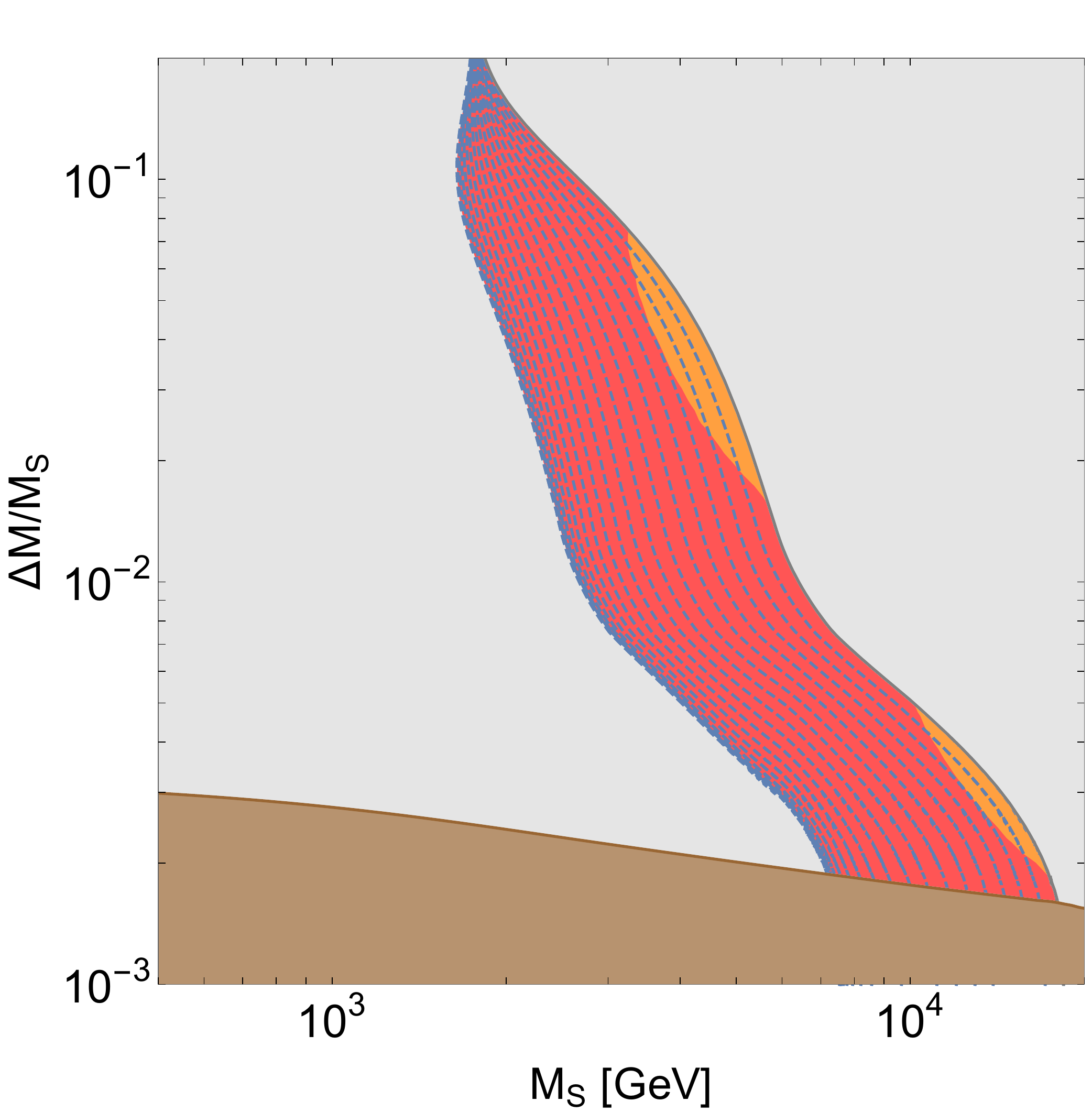}
\caption{Experimental limits on thermally produced real scalar DM coupling via a coloured mediator to $u$-quarks. The left and right panel show the case of the Higgs portal coupling $\lambda_3=0.0$  and $\lambda_3=1.5$, respectively. The null-results of XENON1T exclude the red part of the parameter space. The orange region indicates the reach of a future Darwin-like direct detection experiment. The upper gray area  requires $y\geq2$ in order to reproduce the correct relic abundance. In the lower grey region  the observed relic abundance cannot be achieved via freeze-out since the thermal abundance is set by the gauge interactions of the mediator. In the brown region the mass splitting is so small that the bound states is lighter than twice $M_{S}$. We indicate contours of constant $y$ in the range $\{0.1,2\}$ by the dashed lines. 
\label{fig:light_quarks}
}
\end{figure}

Our results for DM coupling to light quarks are summarized in Fig.~\ref{fig:light_quarks}. The blue dashed lines indicate contours of constant $y$ in the $\Delta M/M_S$ plane. 
The top-most line corresponds to $y=2$ and we show lines down to $y=0.1$ with a $\Delta y=0.1$ spacing.  The region where thermal freeze-out can account for the observed relic density is limited from above and below. 
In the top-region, the effective annihilation cross section is too small to yield the measured DM abundance unless $y>2$ is considered. We prefer not to report results for such large couplings as effects that are higher order in $y$, which are not covered by our calculations, can become relevant in this regime. 
In the lower region the situation is different. The contours of fixed $y$ get closer and closer to each other until they essentially merge at the border of the lower grey region. 
This is due to the fact that for lower $y$ the contributions to $\langle \sigma_{\hbox{\scriptsize eff}} \, v \rangle$  from pure gauge interaction become more important until they completely dominate the annihilation rate. Once this happens the value of $\Delta M/M_S$ required for successful generation of $\Omega_{\hbox{\tiny DM}} h^2$ becomes independent of $y$.\footnote{This statement only holds as long as $y$ is large enough to support efficient conversion between $S$ and $F$. If $y$ is so small that chemical equilibrium between the dark sector states cannot be maintained a new conversion drive-freeze-out mechanism becomes viable~\cite{Garny:2017rxs}.}

As can be seen, the experimental sensitivity is excellent. For small $\lambda_3$, the low-mass large-mass splitting part of the parameter space is probed by LHC searches for multijet+$MET$ (blue region) and $M_F \lesssim 1.5$~ TeV is excluded for large $y$. For smaller values of $y$, the limits weaken since the new physics induced production cross section decreases. For small $y$ we find that  $M_F \geq 600$~GeV is still viable.  In addition, there are strong bounds from direct detection experiments. The current bounds from XENON1T (red region) already exclude parts of the parameter space. Curiously, the sensitivity is best in two disconnected regions at high and low $M_S$. This can be understood from the interplay between the relic density and direct detection. In the low mass region, the experimental sensitivity profits from the larger number density of DM particles and the cross section is larger due to the lower overall mass scale. In the high mass region, the direct detection cross section gets enhanced due to the small mass splitting $\Delta M$ which enters in the tree-level mediator exchange diagrams (see Fig.~\ref{fig:DD}). Therefore, a second exclusion region opens up. In the future, more sensitive direct detection experiments such as DARWIN~\cite{Aalbers:2016jon} have the potential to cover basically all the remaining parameter space. A thin sliver of parameter space in the vicinity of the lower border remains viable since very small values of $y$ are possible there. This region is too small to be visible in the plot.

For large $\lambda_3$ this picture changes in two important ways. First, $SS$ annihilation via the Higgs  becomes efficient and for large $\Delta M /M_S$ we recover the Higgs portal solution for the relic density. This can be seen by the throat-like structure at $M_S \approx 1.7$ TeV into which all relic density lines merge at large $\Delta M /M_S$.  It is interesting to observe that the lines for small $y$ are actually not described by single function and two distinct solutions exist for low masses. This behaviour can be understood by considering the effective annihilation rate defined in Eq.~(\ref{eq:sigmav}). Close to the Higgs portal solution the numerator is dominated by the $\lambda_3$ contribution to $c_1$ which does not depend on $\Delta M /M_S$. As $\Delta M$ decreases the denominator increases while the numerator stays approximately constant. The only way to keep $\langle \sigma_{\hbox{\scriptsize eff}} \, v \rangle$ constant under these conditions is to decrease the mass. For sufficiently small $\Delta M$,  the mediator annihilation piece, which grows faster than the denominator, will become relevant and the usual behaviour of the curves is recovered. Second, the Higgs portal coupling increases the DM-nucleus scattering rate and strengthens the direct detection limits. As  a consequence, the two regions found for $\lambda_3=0$ merge and only a small range of parameters remain consistent with the XENON1T limits. Since the lowest possible DM mass for $\lambda_3=1.5$ is above $1.5$ TeV, the scenario is currently unconstrained by LHC searches.

\begin{figure}[t]
\centering
\includegraphics[width=0.49\textwidth]{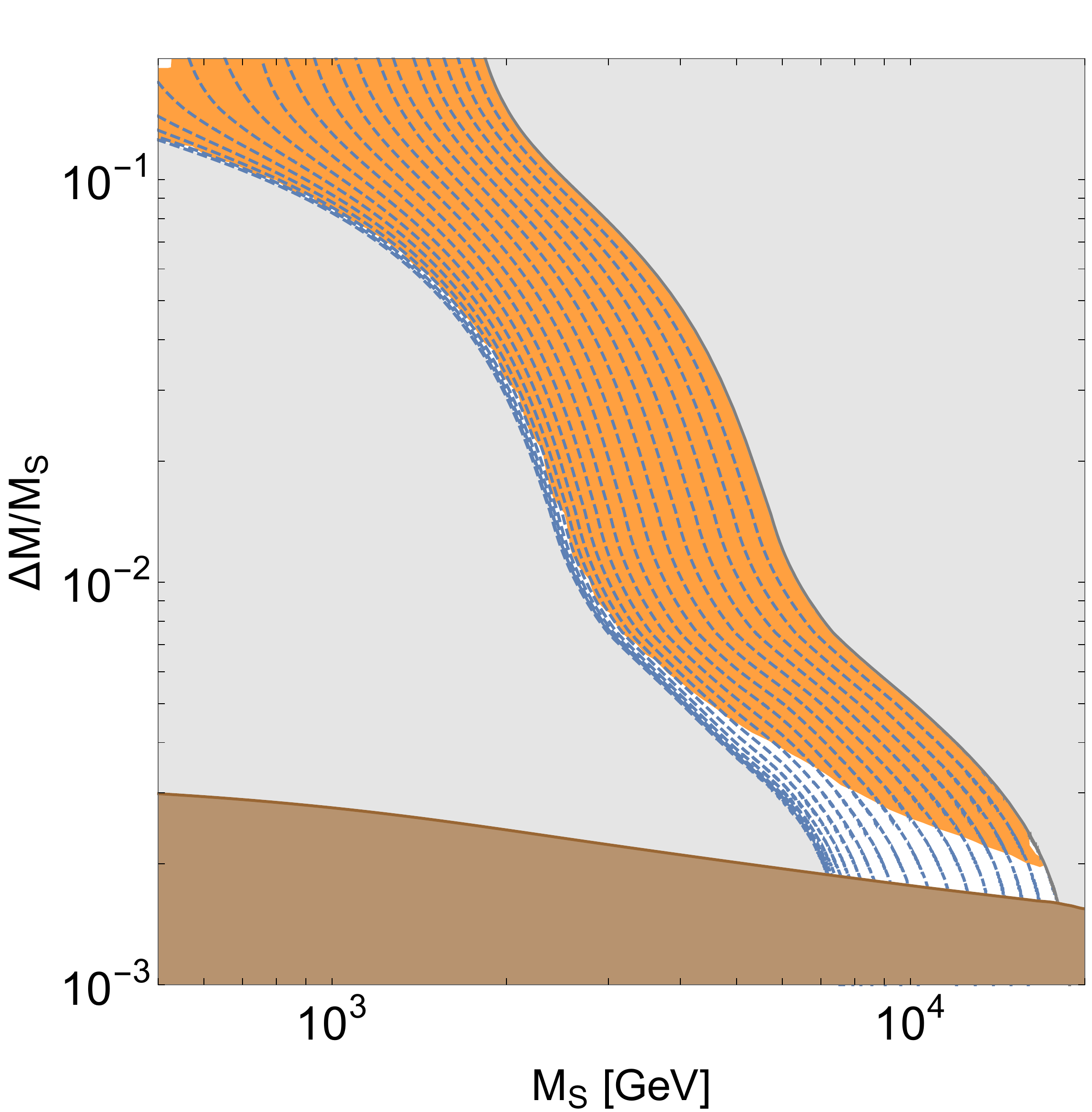}
\includegraphics[width=0.49\textwidth]{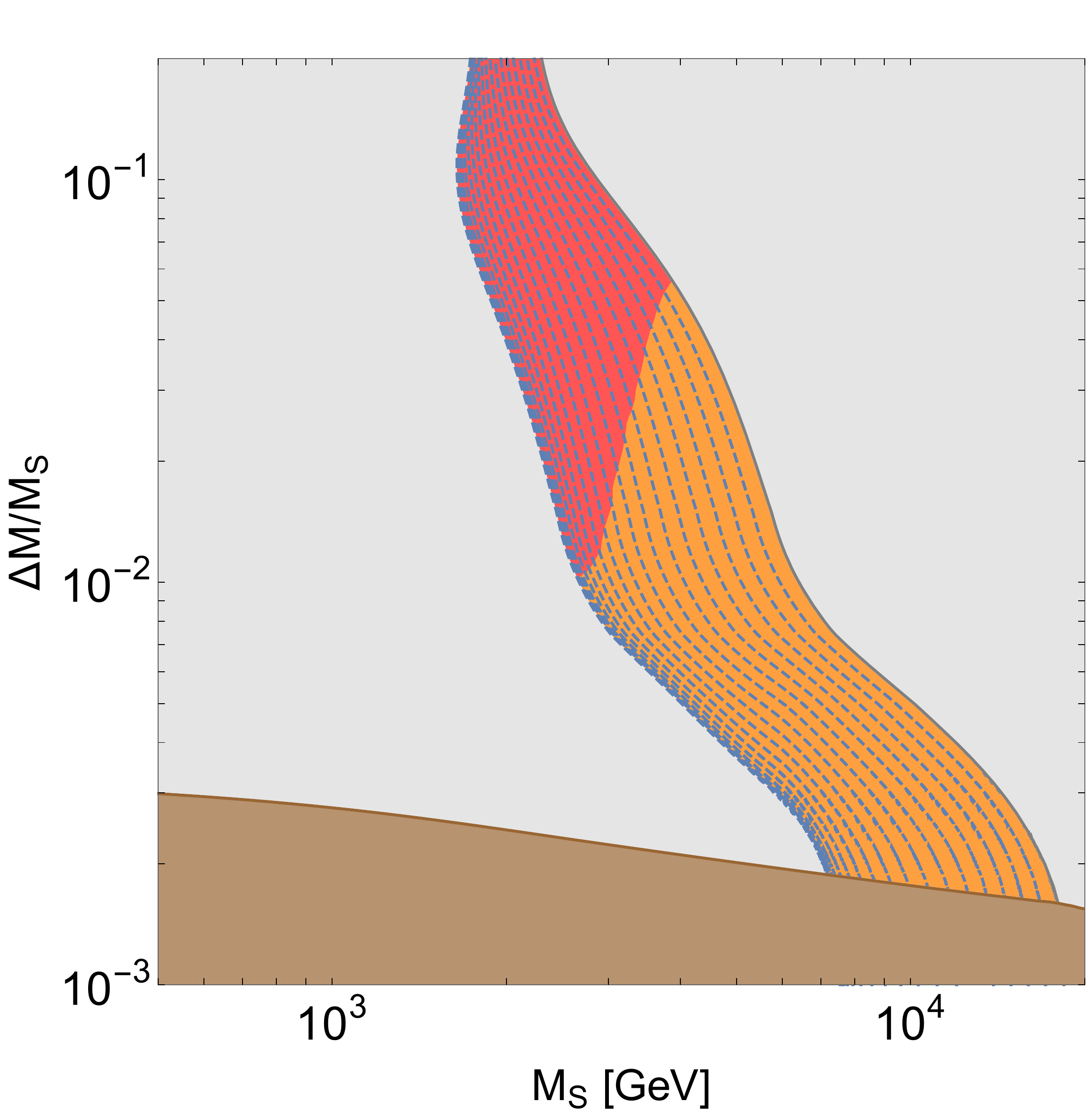}
\caption{Same as Fig.~\ref{fig:light_quarks} for DM coupling to top quarks. Note that $\lambda_3$ always experiences non-negligible running for DM coupling to the top. The value of $\lambda_3(\mu)=0 (1.5)$ has been fixed at the scale of DM annihilations, $\mu=M_S$ in the left (right) panel.
\label{fig:results_top}
}
\end{figure}

The situation changes for DM coupling to top-quarks.
The first difference concerns the relic density.
For $\lambda_3=0$, the most visible difference between the relic density curves in Fig. \ref{fig:light_quarks} and Fig.~\ref{fig:results_top} occur at large $\Delta M /M_S$ and low $M_S$. This is due to the different $SS$ annihilation cross sections. As mentioned previously, the $y$-dependent contribution to $\sigma v (SS\rightarrow q \bar{q})$ scales differently for light quarks and top quarks.  In the first case the lowest order contribution arises at the $d$-wave and the cross section is suppressed by $T^2/M_S^2$. For top quarks on the other hand, the dominant contribution is helicity suppressed, i.e.~scales as $m_t^2/M_S^2$. For light DM $m_t$ is not a small parameter and the suppression of $\sigma v (SS\rightarrow t\bar{t})$ is only mild. Consequently, DM annihilation is more important for top quarks and the parameter space at low masses is more open than in the case of up-quarks. In the high-mass regime, these differences do not matter any more and, for $M_S \gtrsim 3$ TeV, there is no significant difference between the parameter space of the two scenarios.
A more profound difference is revealed when we look at the direct detection cross section. On the one hand, tree-level mediator exchange does not contribute to the direct detection cross section and the only  appreciable $y$-dependent piece arises from the loop-induced coupling to gluons. In the parameter space under consideration here, i.e.~$M_S \geq 500$~GeV the scattering rate from the effective gluon coupling is too small to be observed even with a DARWIN-like detector. However, the running of $\lambda_3$ also has to be treated with greater care now. 
As mentioned previously, the renormalization of the $SSHH^\dagger$ vertex due to box-diagrams with the mediator $F$ and the top-quark in the loop is important.
It leads to the last term in Eq.~(\ref{run_6}), which just depends on $y$ and the SM-Yukawa $h$ and can induce a substantial running even if $\lambda_3$ vanishes. Since $h_t$ is large, this contribution is large for top-philic DM and the solution $\lambda_3(\mu)=0$ is not stable if we vary $\mu$. Since the scale of annihilations ($\mu=M_S$), and the scale of direct detection, i.e the nuclear scale relevant for  direct detection, are very different we have to take the running into account. For $\lambda_3(\mu=M_S)=0$ the phenomenological consequences are quite startling and we find that the Higgs coupling leads to a direct detection rate in reach of DARWIN in most of the relevant parameter.\footnote{It is in principle also possible to require $\lambda_3(\mu)=0$ at the direct detection scale and derive the appropriate value for the relic density calculations. This scenario is by construction not testable with the experiments considered here and we do not discuss it further.} Naturally, the same effect is also relevant for  $\lambda_3(\mu=M_S)=1.5$. However, the impact on the phenomenology is less severe in this case. The main visible consequence in Fig.~\ref{fig:results_top} is the tilt of the red region. In an analysis without running the edge of the red region would have been a straight line. We find that $M_S$ less than $2-3$ TeV is excluded depending on the particular value of $y$. 

All considered, we find good prospects to test the parameter space preferred by thermal freeze-out with experiments in all the scenarios considered above.

%%%%%%%%%%%%%%%%%%%%%%%%%%%%%%%%%%%%%
\section{Conclusions and Outlook}
%%%%%%%%%%%%%%%%%%%%%%%%%%%%%%%%%%%
\label{sec:Conclusions}
In this paper, we address the phenomenology of a simplified model where a scalar DM particle interacts with the SM sector through  a vector-like fermion that induces a Yukawa-type interaction between the DM and a SM quark.  In order to retain gauge invariance, the fermion mediator gauge quantum numbers have to match those of the quarks. In addition, scalar DM can  interact with the Standard Model via the Higgs portal.
This model implementation opens up a rich phenomenology while the minimal Higgs-portal setting can be recovered for the limiting case $y \to 0$. 

We mainly focus on the coannihilation regime, namely when the additional mediator $F$ is close in mass with the DM particle. In this case, one has to carefully track the dynamics of the accompanying coloured state to determine the DM abundance. This is a challenging task since  Sommerfeld enhancement and bound-state effects in a thermal medium can drastically change the annihilation cross section for mediators with QCD interactions. We use a non-relativistic effective field theory framework to relate the annihilation cross sections to the spectral functions of non-relativistic particle pairs. The spectral functions are extracted from the solution of a plasma-modified Schr\"odinger equation which takes the thermal potentials, the collisional damping rate and the dissociation rates as input parameters. With respect to previous results, where the attention was on the above threshold states, we include the bound-state effects for both the singlet and antitriplet colour configuration of $F\bar{F}$ and $FF$ pairs. Together with the unbound octet and sextet states, they are recast in the language of pNRQCD. 
In doing so, we provide an accurate description of the coloured pairs in the early Universe which allows the reliable extraction of the relic density. The main outcome is that enhanced cross sections, which are driven by the strongly interacting mediator (co)annihilations, require quite large DM masses to reproduce the observed relic abundance.

We would like to comment briefly on the form of the Boltzmann equation (\ref{BE_gen}) that we adopt in this work. This quadratic equation in the dark particles number density appears to break down when the temperature is much smaller than the binding energy, and  an alternative form of the rate equation has recently been suggested in ref.~\cite{Binder:2018znk}. In our analysis, this limitation is taken into account by the brown band appearing in Fig.~\ref{fig:light_quarks} and \ref{fig:results_top}, that implements a lower bound on the mass splitting $\, 2 \Delta M >  |E_1'|$, where $E_1'$ is the binding energy of the lowest-lying bound state. The corresponding brown region is where the relic density extraction from the standard Boltzmann equation for coannihilation could be inaccurate.  A recent study~\cite{Biondini:2019zdo} suggests that the in-vacuum mass splitting is the appropriate regulator even if the ameliorated rate equation in ref.~\cite{Binder:2018znk} is considered. Therefore, it seems necessary to better understand how to switch from a single rate equation to a set of coupled
equations for the scattering and bound states  at very low temperatures.

We assess the capabilities of the different experimental strategies to probe the parameter space compatible with the observed relic density. For the high-mass range we focus on, $M_S \geq 500$ GeV, collider searches and direct detection are the relevant options. The experimental signatures depend crucially on the type of SM quark the DM interacts with. We consider two extreme cases, valence quarks represented by the up quark and the top quark. 
In general, the LHC phenomenology of coannihilation scenarios is characterized by comparatively soft SM particles in the final state. This leads to a loss of sensitivity of the searches compared to scenarios with an unsuppressed mass splitting between the DM and the mediator.    
Nevertheless, for couplings to up-quarks, the LHC can probe the low-mass  large-$y$ part of the parameter space. This is due to the enhanced production of the mediator from $t$-channel DM exchange, which can dominate the production rate for $y\gtrsim 0.7$. 
This mechanism is not active for DM interacting with the top and the LHC is currently not able to test this scenario. 
Fortunately, direct detection experiments have a good sensitivity to the coannihilation region.
For couplings to up-quarks a substantial part of the parameter space is 
already excluded by the null searches of XENON1T irrespectively of the Higgs portal coupling. The unconstrained regions are well within the reach of a Darwin-like detector even though freeze-out allows very large DM masses of up to $18$ TeV.
The case of direct detection for DM coupling to tops is more subtle. As the tree-level interaction vanishes and the effective interaction with the gluons is highly suppressed, the only potentially detectable contribution to DM nucleus scattering arises from the Higgs portal term. As $\lambda_3$ is a free parameter of our theory, one might come to the conclusions that one can simply set this term to zero and avoid all direct detection constraints. While this is true in principle, it has to be kept in mind that the scale of direct detection and the scale of annihilations are rather different and, therefore, 
the running of the coupling can be important. We investigate the renormalization of $\lambda_3$ in a non-relativistic theory for the DM and the mediator.
As $\Delta M \ll M_S$ the mass splitting remains a dynamical scale in this theory and box-diagrams with the $F$ and the top quark in the loop contribute to the renormalization of the $SS H H^\dagger$ vertex. This results in a contribution to the corresponding $\beta$ function that does not depend on $\lambda_3$, but rather on the Yukawa couplings $y$ and $h_q$. Therefore, a large SM Yukawa $h_q$ generically lead to substantial running and
we find that it is not possible to set $\lambda_3=0$ at both scales simultaneously. 
This has consequences for the phenomenology and we find that future direct detection experiments are sensitive to most of the parameter space if $\lambda_3(\mu=M_S)=0$, whereas for  $\lambda_3(\mu=M_S)=1.5$ some parts of the cosmologically preferred parameter space are already excluded.

\section*{Acknowledgements}
S.B. thanks N. Brambilla, M. Escobedo, M. Laine, V.~Shtabovenko and A. Vairo for helpful and valuable discussions.

\appendix
\numberwithin{equation}{section}

\setcounter{equation}{0}

%%%%%%%%%%%%%%%%%%%%%%%%%%%%%%%%%%%%%%%%%%%%%%%%%%%%%%%%%%%%%%%%%%%%%%%%%%%%%%%%%%%%%%%%%%%%%%%%%%%%%%%%%%%%%%%%%%%%%%%%%%%
\section{Thermal potentials and numerical implementation}
\label{App_pNRQCD_1}
In this appendix, we provide some details for the derivation of Eqs.~(\ref{result1_mpi}) and (\ref{result2_mpi}), together with comments on the numerical strategy and the running of $\alpha_s$.

The thermal corrections to the potential may be obtained by calculating loop corrections to the longitudinal gluon polarization tensor. There are two different possibilities for the particles running in the loop. They can be either fermions (quarks) or gluons. Moreover, there is a real and imaginary part in both cases. In the following, we focus on the real part since the imaginary part in the regime under study is known \cite{Brambilla:2013dpa}. 

The starting point is the one-loop thermal self-energy of the gluon\cite{Kajantie:1982xx}. We are interested in the typical loop momentum to be of the order of the temperature, which we impose to be of the same order of the spatial  momentum carried by the gluon $\bm{p}\equiv p \sim 1/r$, namely the momentum transfer between the two non-relativistic fermions (and the inverse size of the heavy pair). Expanding in the energy transfer $p_0 \approx p^2/(2M_F) $ which is much smaller than $p$, the real part of the retarded gluon self-energy reads~\cite{Kajantie:1982xx,Heinz:1986kz,Brambilla:2008cx}
\begin{eqnarray}
&&{\rm{Re}} \left[ \Pi^{R}_{00}(p)\right] = -\frac{g^2 T_F N_f }{\pi^2} \int^{\infty}_{0} dk_{0} \, k_0 \, n_F(k_0) \, \left[ 2+\left( \frac{p}{2k_0} -2\frac{k_0}{p}\right) \ln \frac{|p-2k_0|}{|p+2k_0|}  \right]  \nonumber \\
&& \phantom{xx}- \frac{g^2  N_c }{\pi^2}  \int^{\infty}_{0} dk_{0} \, k_0 \, n_B(k_0) \, \left[ 1- \frac{p^2}{2k^2_0} + \left( -\frac{k_0}{p} + \frac{p}{2k_0} - \frac{p^3}{8k^3_0}\right)  \ln \frac{|p-2k_0|}{|p+2k_0|}  \right] \, ,
\label{EQ2}
\end{eqnarray}
where we keep terms up to order $p_0$.  Moreover, $n_F(x)=1/(e^{x/T}+1)$ and $n_B(x)=1/(e^{x/T}-1)$ are the usual Fermi-Dirac and Bose-Einstein distributions.

We work in the real-time formalism and in order to obtain the corrections to the $D_{11}$ gluon propagator, namely the self-energy corresponding to the positive branch of the Keldysh contour, we follow the same procedure as in \cite{Escobedo:2010tu}. The following relations are needed:
\begin{eqnarray}
&&D_{11}=\frac{1}{2}\left( D_R + D_A + D_S \right) \,  , \nonumber \\
&&D_R=D^{0}_R + D_R \Pi_R D^{0}_R \, ,
\label{Dys_R}
\end{eqnarray} 
where $D_R$, $D_A$ and $D_S$ are the retarded, advanced and symmetric gluon propagators, respectively. It is convenient to work with the retarded propagator and self-energy because the Dyson equation (\ref{Dys_R}) is of zero-temperature type.
Finally, we use the following definition to relate the propagator and the potential 
\begin{equation}
V(r)=-C_Fg^2 \mu^{4-D} \int \frac{d^{D-1}p}{(2 \pi)^{D-1}} \left( e^{i\bm{p}\cdot \bm{r}}-1\right) D_{11}(p_0=0,p) \, ,
\label{EQ3} 
\end{equation}
where we have included also the self-energy contribution for the heavy quark and anti-quark pair. Working out the integral in dimensional regularization, the final expressions in Eqs.~(\ref{result1_mpi}) and (\ref{result2_mpi}) are found.

We comment briefly on the setting for the numerical evaluation of the plasma-modified Schr\"odinger equation; for a more detailed discussion we refer the reader to Refs.\cite{Biondini:2018pwp,Biondini:2018ovz}. 
When evaluating the static potentials, a wide range of distance scales appears in the Schr\"odinger equation. At short
distances, we evaluate the 2-loop coupling at the scale $ \bar{\mu} = e^{-\gamma_{\hbox{\tiny E}}}/r$ \cite{Schroder:1998vy,Lee:2016cgz}.
Since parametrically only the scales $ \alpha_s M_F \ll M_F $ play a role in the Schr\"odinger equation, the
running does not include the coloured fermion $F$ in this domain. Instead, at the scale of the hard annihilation in the early Universe, we have taken a one-loop running for the strong coupling with the additional fermion included, that reads
\begin{equation}
\partial_t g_s^2 = \frac{g_s^4}{(4 \pi)^2} \Bigg\{  \frac{4N_G}{3} + \frac{2N_F}{3}  -\frac{11}{3}N_c \Bigg\} \, \, ,
\end{equation}
where $t = \ln \bar{\mu}^2$, $N_G= 3$ is the number of SM generations and $N_F$ is the number of additional fermion mediators ($N_F=1$ in our case). 

At large distances, we use effective thermal couplings that are taken from a dimensionally reduced field theory \cite{Ginsparg:1980ef,Appelquist:1981vg}. The Debye mass $m_D$ and the
electrostatic coupling are derived in \cite{Laine:2006cp}. We use their expressions that take into account the SM quark mass thresholds at different temperatures, so to implement a varying number of SM quark flavours $N_f$ entering the running.

%As far as the thermal potentials for the gluodissociation are concerned, we impose them to contribute when $1/r > \pi T$ is
%satisfied. This is needed to justify the usage of that potential in first place, and also to avoid
%spurious effects in the spectral function determination at large distances.
%%%%%%%%%%%%%%%%%%%%%%%%%%%%%%%%%%%%%%%%%%%%%%%%%%%%%%%%%

\section{NREFT and renormalization group equations}
\label{app:RGE}

In this appendix, we describe the effective field theory  for non-relativistic DM particles and the accompanying coloured fermion. Such a low-energy theory can be derived from the fundamental theory in (\ref{Lag_RT}) by removing energy/momenta of order $M_S$, i.e.~when implementing a $1/M_S$ expansion in Eq.~(\ref{Lag_RT}). The non-relativistic degrees of freedom we consider are the DM scalar and the coloured fermion. This situation is realized when the mass splitting is small with respect to the DM mass, which is the coannihilating framework we considered in this work.\footnote{This situation is different from the case of a very heavy mediator that is integrated out first
~\cite{DEramo:2014nmf}.} As the mass splitting is still dynamical in the low-energy theory, the coloured mediator has a small residual mass term in the corresponding non-relativistic propagator. A similar EFT has been addressed for nearly degenerate Majorana neutrinos in ref.\cite{Biondini:2015gyw}. In addition, the effective theory contains the SM sector, which is assumed to be in the unbroken phase.

Our aim is to address the running of the couplings from the DM mass scale down to the mass splitting/electroweak breaking scale, which are comparable over the parameter space that reproduces the relic abundance. The running can be important for direct detection cross section which crucially depends on the effective vertex between two DM particles and two Higgs fields. In the NREFT, this interaction corresponds to an operator of dimension five and, therefore, we will only study operators up to this order.

As far as the coloured mediator is concerned, the prototype of the low-energy theory is HQEFT\cite{Neubert:1993mb}, that has to be supplemented with a residual mass $\Delta M$. In a given reference frame, the momentum of a non-relativistic DM or coloured fermion field is $M_S v^\mu$, with $v^2= 1$, up to fluctuations whose momenta $k^\mu$ are much smaller than $M_S$  (we can indeed take the same velocity because of the small mass splitting).
Up to dimension-5 operators, the EFT reads 
\begin{eqnarray}
    \mathcal{L}_{\hbox{\tiny NREFT}}&=&\mathcal{L}_{\hbox{\tiny SM}}+\phi^\dagger \left( v \cdot \partial -\frac{D_\perp^2}{2 M_S} \right) \phi 
    \nonumber \\
    &+& \bar{\psi} \left( i v \cdot D -\Delta M - \frac{D_\perp^2}{2 M_S} -\frac{g_s \, \sigma_{\alpha \beta} F^{\alpha \beta}}{4 M_S} \right) \psi + (\psi \to \chi)
    \nonumber \\
    &-& \left( \frac{y}{\sqrt{2 M_S}} \, \phi  \, \bar{\psi} \, P_R \, q  + \frac{y}{\sqrt{2 M_S}} \, \phi^\dagger \,  \bar{\chi} \, P_R \, q  + {\rm{h.c.}} \right) - \frac{\lambda_3}{2 M_S} \phi^\dagger \phi H^\dagger H %+ \frac{c_\psi}{M} \bar{\psi}\ps \, ,i H^\dagger H 
    + \cdots \, ,
    \nonumber 
   \\
   \phantom{s}
    \label{non-rel_appendix}
\end{eqnarray} 
where the dots stand for operators that are further suppressed by $1/M_S$, $D^\mu=\partial^\mu + i g_s A^{\mu \, a} T^a$ and $\partial^\mu_\perp=\partial^\mu-v \cdot \partial$, $\sigma_{\alpha \beta}=i[\gamma_\alpha, \gamma_\beta]/2$. The matching from the relativistic theory (\ref{Lag_RT}) and the EFT (\ref{non-rel_appendix}) has been done at tree level. We should write an operator $\bar{\psi}\psi H^\dagger H $, however there is no contribution to it at tree level from the fundamental theory (\ref{Lag_RT}).  A comment is in order for the scalar non-relativistic field in Eq.~(\ref{non-rel_appendix}). Upon the field redefinition $\phi \to \phi / \sqrt{2M_S}$, we obtain a non-relativistic scalar field with mass dimension $[\phi]=3/2$ just as for the fermion field $\psi$ (and $\chi$). This is also the choice made in Sec.~\ref{sec:Relic_density} when writing the four-particle operators for annihilations~\cite{Biondini:2017ufr,Biondini:2018pwp}. The scalar propagator takes the same form as the fermion one in Eq.~\ref{non-rel_appendix}. We note that the operator comprising the Yukawa coupling $y$ has dimension $9/2$. 
The DM self-interaction in (\ref{Lag_RT}), that involves the coupling $\lambda_2$, is a dimension-6 operator in the $1/M_S$ expansion. Therefore we neglect it as far as the running is concerned.

Since we want to obtain the one-loop running of the $y$ and $\lambda_3$ couplings relevant for the direct detection, we need to extract the counterterms at one-loop for the Lagrangian (\ref{non-rel_appendix}). An EFT can be renormalizable systematically order by order in the large scale expansion. In our case, we only look at the counterterms that renormalize dimension-5 operators at most. Therefore, we do not account for diagrams that induce divergencies that would need counterterms from higher dimensional operators. The power counting of the vertices guide us to organize the set of diagrams relevant for the one-loop computation, that is performed in the rest frame of the heavy fields. 
\begin{figure}
    \centering
    \includegraphics[scale=0.50]{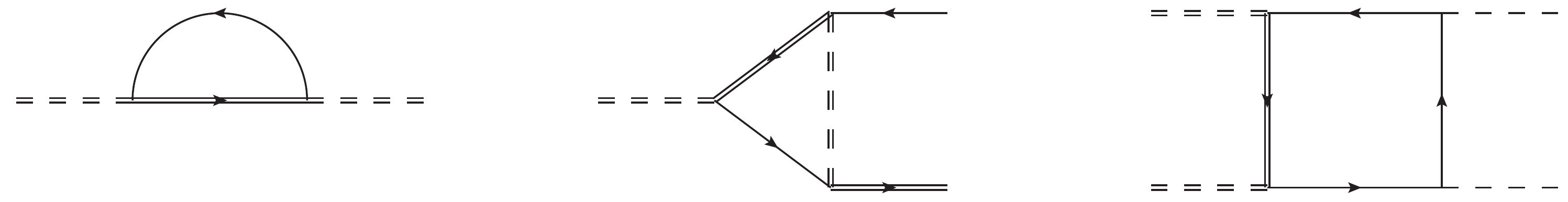}
    \caption{Examples of one-loop diagrams relevant for the running couplings $y$ and $\lambda_3$. From left to right we have displayed the self-energy diagram for the scalar DM, the Yukawa triangle vertex and the box-diagram. Dashed double lines stand for the DM, solid double line for the mediator, then solid lines for the SM right-handed quarks. In the latter diagram, the thin solid lines stand for the right-handed top and left-handed quark doublet attached to the Higgs boson external legs. }
    \label{fig:EFT_SS}
\end{figure}

We reproduce the known result for the wave renormalization of $\psi$ induced by QCD interactions, which is the same as for an heavy quark in QCD\cite{Politzer:1988wp}. In addition, we find a suppressed $\Delta M/M_S$ contribution to $\psi$ field renormalization induced by the Yukawa vertex. The QCD vertex involving a magnetic gluon is not renormalized\footnote{One has to notice that this would correspond to a transition between the same heavy coloured state, then fixing the function in ref.\cite{Falk:1990yz} to be evaluated for $v'=v$. Together with the wave function renormalization from QCD, the vertex does not produce a divergence.}.
We list the equations that describe the running couplings for $m_W, \Delta M \leq \bar{\mu} \leq M_S$. Using dimensional regularization with $D=4-\varepsilon$  in the $\overline{\text{MS}}$ scheme, and $t = \ln \bar{\mu}^2$, we find 
\begin{eqnarray}
&&\partial_t \mu_H^2 = \frac{1}{(4 \pi)^2} \Bigg\{ \left[ 6 \lambda_1 -\frac{9}{4} g_w^2 + |h|^2 N_c\right] \mu_H^2 \Bigg\} \, , 
\label{run_1}
\\
&&\partial_t g_s^2 = \frac{g_s^4}{(4 \pi)^2} \Bigg\{  \frac{4n_G}{3}   -\frac{11}{3}N_c \Bigg\} \, , 
\label{run_2}
\\
&&\partial_t |y|^2 = \frac{|y|^2}{(4 \pi)^2} \Bigg\{ -4 \frac{\Delta M}{M_S} |y|^2 (1+  N_c) + |h|^2 -3 g_s^2 C_F \Bigg\} \, , 
%\label{run_3}
\\
&&\partial_t |h|^2 = \frac{|h|^2}{(4 \pi)^2} \Bigg\{ |h|^2 \frac{(2 N_c +3)}{2}  -\frac{9g_w^2}{4}-6 g_s^2 C_F \Bigg\} \, , 
\label{run_4}
\\
&&\partial_t \lambda_1=\frac{1}{(4 \pi)^2} \Bigg\{ \left[ 12 \lambda_1 -\frac{9}{2} g_w^2 +2 |h|^2 N_c\right]\lambda_1 + \frac{9g_w^4}{16}-|h|^4 N_c \Bigg\} \, , \label{run_5}
\\
&&\partial_t \lambda_3=\frac{1}{(4 \pi)^2}  \Bigg\{ \left[ 6 \lambda_1 - \frac{9 g_w^2}{4} -4  N_c |y|^2 \frac{\Delta M}{M_S} + N_c |h|^2 \right] \lambda_3 
-2  N_c |y|^2 |h|^2   \Bigg\} \, .
%\label{run_6}.
\end{eqnarray}
It is perhaps useful to comment on the $\Delta M / M_S$ suppression that one finds in Eqs.~(\ref{run_3}) and (\ref{run_6}). This originates from the one-loop DM self-energy diagram and the one-loop vertex diagram, see Fig.~\ref{fig:EFT_SS}. As far as the first diagram is concerned, the $1/M_S$ suppression, coming from the insertions of two tree-level Yukawa vertices, is balanced by the dynamical scale $\Delta M$, that runs in the loop because of the coloured mediator, and that multiplies the $1/\epsilon$ pole of the corresponding UV divergencies. As a result, this contribution match the dimension-4 required for the $\phi^\dagger (v \cdot \partial) \phi$ operator. A similar situation occurs for the vertex diagram, keeping in mind that the additional $M_S^{-1/2}$ suppression is balanced from that present in the tree-level vertex.  Instead, the box diagram that contributes to the dimension-5 operator $\phi^\dagger \phi H^\dagger H$, diagram on the left in Fig.~\ref{fig:EFT_SS}, displays  the correct $1/M_S$ suppression from the insertion of the vertices, and we find a divergent term of the form $1/\varepsilon$ without any accompanying energy scale. 

\bibliographystyle{hieeetr}
\bibliography{Scalar_DM.bib}

\end{document}